\documentclass[12pt,a4paper]{article}
\newlength{\overeqskip}
\newlength{\undereqskip}
\usepackage{amssymb}
\usepackage[pdftex]{graphicx}

\def\nc{\newcommand}
\nc{\half}{\frac{1}{2}}
\nc{\shalf}{\ensuremath{\textstyle \frac{1}{2}}}

\nc{\deldag}{\mathbin{\partial\mkern-10.5mu\big/}}
\nc{\deldagss}{\mathbin{\partial\mkern-10.5mu/}}
\nc{\kdag}{\mathbin{k\mkern-10mu\big/}}
\nc{\udag}{\mathbin{u\mkern-10mu\big/}}
\nc{\kdagss}{\mathbin{k\mkern-10mu/}}
\nc{\Pdag}{\mathbin{P\mkern-10mu\big/}}
\nc{\pp}{{\scriptscriptstyle ||}}
\nc{\stwo}{{\scriptscriptstyle 2}}
\nc{\pham}{{\phantom{-}}}

\def\lsim{\mathrel{\raise.3ex\hbox{$<$\kern-.75em\lower1ex\hbox{$\sim$}}}}
\def\gsim{\mathrel{\raise.3ex\hbox{$>$\kern-.75em\lower1ex\hbox{$\sim$}}}}

\def\Slashnew#1{#1\kern-0.55em\raise.05ex\hbox{/}}
\def\slashnew#1{#1\kern-0.5em\raise.05ex\hbox{{$\scriptstyle /$}}}
\def\ie{{\em i.e. }}
\def\wrt{{\em w.r.t. }}
\def\eg{{\em e.g. }}
\def\hi{\phantom{i}}

\def\etal{{\it et.al.~}}

\nc{\beq} {\begin{equation}}
\nc{\eeq} {\end{equation}}
\nc{\beqa}{\begin{eqnarray}}
\nc{\eeqa}{\end{eqnarray}}
\def\emph#1{{\em #1}}
\def\hepph#1{hep-ph/#1}
\def\hepth#1{hep-th/#1}

\begin{document}
%
%
\begin{titlepage}
\pagestyle{empty}
\baselineskip=21pt
\vskip .6in

\begin{center} {\Large{\bf Towards a kinetic theory for fermions with quantum coherence}}

\end{center}
\vskip .3in

\begin{center}

Matti Herranen,  Kimmo Kainulainen and Pyry Matti Rahkila\\

\vskip .2in

{\it  University of Jyv\"askyl\"a, Department of Physics, P.O.~Box 35 (YFL),
        \\ FIN-40014 University of Jyv\"askyl\"a, Finland }\\
        and \\
\vskip .1in
{\it  Helsinki Institute of Physics, P.O.~Box 64, FIN-00014 University of  		
   	       Helsinki, Finland.}\\

\end{center}

\vskip 0.3in

\centerline{ {\bf Abstract} }
\baselineskip=16pt
\vskip 0.5truecm\noindent
\\
A new density matrix and corresponding quantum kinetic equations are introduced for fermions undergoing coherent evolution either in time (coherent particle production) or in space (quantum reflection). A central element in our derivation is finding new spectral solutions for the 2-point Green's functions written in the Wigner representation, that are carrying the information of the quantum coherence. Physically observable density matrix is then defined from the bare singular 2-point function by convoluting it with the extrenous information about the state of the system. The formalism is shown to reproduce familiar results from the Dirac equation approach, like Klein problem and nonlocal reflection from a mass wall. The notion of the particle number in the presence of quantum coherence is shown to be particularily transparent in the current picture. We extend the formalism to the case of mixing fields and show how the usual flavour mixing and oscillation of neutrinos emerges again from a singular shell structure. Finally, we show how the formalism can be extended to include decohering interactions. 

\vskip 2truecm
\noindent matherr@phys.jyu.fi, kainulai@phys.jyu.fi, pmrahkil@phys.jyu.fi

\end{titlepage}

\baselineskip=17pt

%

\section{Introduction}

Many problems in modern particle physics and cosmology require setting up transport equations for quantum systems in out-of-equilibrium conditions, including electroweak baryogenesis~\cite{EWBGrew}, leptogenesis~\cite{buchmuller} and particle creation in the early universe~\cite{TomGar}, just to mention a few.  Moreover, one often should be able to describe the evolution of coherent quantum correlations in the presence of decohering effects of the surroundings. This is true in particular for the case of the electroweak baryogensis (EWBG), where the problem is to reliably compute the fermionic out-of-equilibrium distribution functions in the neigbourhood of an expanding CP-violating phase transition wall.
The EWBG problem can be divided to two regimes depending on the width of the phase transition wall in comparison to the mean free path of the fermions interacting with the wall. The case of thick wall has been addressed earlier in the semiclassical WKB approach in~\cite{CJK,CJK2} (for earlier work see also~\cite{Earlier}) and later in the context of CTP formalism in~\cite{KPSW,gauge}. For a review see~\cite{PSW}. In the thick wall limit the notion of localized particle distributions can be maintained and one can reduce the full quantum transport equations to local Boltzmann equations involving  CP-violating (and CP-even) force terms employing a well defined expansion in spatial gradients. In the thin wall limit the dominant source for the asymmetry comes from the quantum reflection processes which are inherently nonlocal and no consistent quantum field theoretical formalism exists for treating reflection phenomena together with decohering collisions. For early attempts to include collisions in a Dirac equation approach see~\cite{hernandez,splicewall,rius}. One of the goals of this paper is to derive from field theory a density matrix formalism that can be used to solve this problem consistently. However, our methods can equally well be used to describe for exsample the coherent particle production or neutrino-oscillations in the early universe. 
Although our motivation comes mostly from cosmological applications and we will only consider fermionic fields, the formalism that we will develop is not restricted to solving only these problems. Instead, our generic approach to coherence within quantum field theory should be easily extended to the case of scalar fields and also to nonrelativistic problems. It should have a wide range of applications in generic problems where one is interested in quantitative description of quantum coherence in noisy backgrounds. 

In this paper we shall consider only the noninteracting problem, but the formalism we develop is easily extendable to the case with interactions. Our main result is finding a phase space description for the quantum coherence in terms of new singular solutions in close analogy with the usual on-shell particle distributions. The basic objects of our study are the 2-point {\em Wightmann} functions\footnote{Note that our function $G^<$ does not contain an explicit minus sign often included to its definition in the literature.}:
\beqa
  iG^<_{\alpha\beta ,ij}(u,v) &\equiv&
         \langle  \bar \psi_{\beta ,j}(v){\psi}_{\alpha ,i}(u)\rangle
      \equiv 
       {\rm Tr}\{\hat \rho \, \bar \psi_{\beta ,j}(v){\psi}_{\alpha ,i}(u) \}
\nonumber \\
  iG^>_{\alpha\beta ,ij}(u,v) &\equiv&
         \langle  {\psi}_{\alpha ,i}(u)\bar \psi_{\beta ,j}(v)\rangle
     \equiv 
      {\rm Tr}\{\hat \rho \, {\psi}_{\alpha ,i}(u) \bar \psi_{\beta ,j}(v)\} \,,
\label{G-less0}
\eeqa
which describe the most interesting properties of the out-of-equilibrium fermionic system. In section \ref{propagatorth} we will first derive the standard form of these functions under the usual assumption of translational invariance both in space and time and in thermal equilibrium. We then generalize this derivation to the case where the translational invariance is lost either in time or in one of the spatial directions (denoted by $z$ hereafter). These studies are most easily done in a mixed representation, where the functions (\ref{G-less0}) are Fourier transformed with respect to the relative coordinate $u-v$. In this representation the equation of motion for $G^<$ is found to separate into two sets of equations we call {\em kinetic}, or evolution equations containing explicit space or time derivatives, and to algebraic (in the mean field limit) {\em constraint} equations. How this division takes place depends on the special assumption on the spacetime symmetries.

Our most important observation is that giving up the translational invariance allows new solutions in the {\em dynamical} phase space that carry information on the quantum coherence in the system. In the time-dependent, but spatially homogenous case considered in section \ref{sect:tcase} this new class of solution is found at shell $k_0 = 0$ and in the planar symmetric static case, studied in section \ref{sect:zcase}, at shell $k_z = 0$. (For a stationary problem, the position of the latter shell is shifted to $k_z = v_wk_0$, where $v_w$ is the velocity of the static frame.) These solutions are interpreted to describe the coherence between particles and antiparticles of opposite 3-momentas and same helicities on mass shells $k_0 =\pm \sqrt{\vec k^2 + m^2}$ in the former case and between left and right moving states of same spin on shells $k_z =\pm \sqrt{k_0^2 - m^2}$ in the latter case. The new coherence solutions are present only in the dynamical functions $G^<$ and $G^>$. We will show that the spectral sum rule excludes these solutions from the spectral function ${\cal A} = i(G^>+G^<)/2$, so that the coherence shells are not a part of the {\em kinematical} phase space. This is as it should be, since an asymptotic state made out of pure coherence without the mixing mass-shell states does not make any physical sense. Moreover, the coherence solutions in $G^{<,>}$ are shown to be inconsistent both with the full translational invariance and with thermal equilibrium.  

The phase space shell structure described above is entirely set by the algebraic constraint equations. At first sight this singular structure appears to render the kinetic equations to be of little use, but in the end the problem reveals an interesting connection to the measurement theory. In section \ref{sect:dynamicaleqns} we show how physical, observable density matrix can be defined as a convolution of the singular phase space solution with a smooth phase space weight function that describes the existing external information on the system. This definition of physical density matrices is the second major result in this paper, along with our finding of the singular coherence shells in the phase space, and we will illustrate this principle with several examples. We show, for example, how a complete information of the momentum, the energy and spin of the state renders the quantum evolution to a trivial constant propagation of an eigenstate without coherence. Other nontrivial examples with quantum mixing include the usual Klein problem, reflection off a smooth phase transition wall and the definition and evolution of the particle number in a homogenous out-of-equlibrium system. In all these cases we are able to reproduce the known results in a way which underlines the appropriate choice of the weight function and the necessity of including the new coherence solutions.

In section \ref{sect:flavourmixing} we extend our formalism to the case of mixing fields and show how the usual notion of the flavour mixing arises in the present context. Introducing flavour mixing through a Hermitian $N\times N$ mass matrix leads to a very complicated shell structure in the phase space. Assuming a suffiently large $k_0$ and no mass-degeneracies, each mass shell solution separates into $N$ separate diagonal and $N(N-1)$ off-diagonal mass shells. Similarly, the number of free coherence functions at shell $k_z=0$ (in the static planar symmetric case) multiplies to $N$ and in addition $N(N-1)$ new coherence shells appear near, but not exactly at $k_z=0$. The role of each shell in the mixing phenomenon is described qualitatively. We show in particular how the usual density matrix equation for (flavour) mixing neutrino system emerges through a use of a weight function that encodes enough information to tell the direction of motion of the state, but not enough to collapse it into a singular mass-eigenstate. In section \ref{sect:interaction} we outline how our formalism can be straightforwardly extended to the case with interations and finally, section \ref{sect:conclusions} contains our conclusions and outlook.

\section{Propagator theory}
\label{propagatorth}

In this section we will first review the standard derivation of a free  fermion propagator in the thermal field theory, concentrating on the role of the underlying assumptions of translational invariance. We then extend the analysis to the case where the mass of the field can depend on the space and time coordinates. The loss of translational invariance leads to a rich structure of solutions for the 2-point function, and in particular to an emergence of quantum coherence as will be seen in sections \ref{subsec:t-shells} and \ref{subsec:z-shells}. To be specific, we suppose that the Lagrangian of the theory is given by
\beq
{\cal L}_{\rm free} = i\bar \psi \deldag \psi
                    + \bar \psi_L m \psi_R
                    + \bar \psi_R m^* \psi_L \,,
\label{freeLag1}
\eeq
where the mass $m = m(t,{\bf x})$ can be complex. This convention follows from the electroweak baryogenesis application where effective masses arising from higgs mechanism are spatially and temporally varying near the first order phase transition fronts and the complex mass is needed for CP-violation. Temporally varying mass term can arise for example in the context of particle prodution in the early universe~\cite{TomGar}.  For such nontrivial mass functions the Lagrangian ${\cal L}_{\rm free}$ can be understood as an effective theory for a fermion in a temporally and spatially varying background field. 

\subsection{Standard derivation of a thermal propagator}
\label{sect:standardCTPprop}

Let us first consider the case where $m$ is constant. In the standard approach to thermal field theory in real time formalism, one introduces a complex time argument defined on some complex time path, an example of which is shown in figure \ref{fig:keldysphath}, and introduces a propagator with complex time ordering 
\beq
G_{\cal C}(u,v) = \theta_{\cal C}(u_0-v_0) G^>(u,v) 
                - \theta_{\cal C}(u_0-v_0) G^<(u,v) \,,
\eeq
which, in the absence of interactions, obeys the equation:
\beq
(i\deldag_u - m^*P_L - mP_R) G_{\cal C}(u,v)  = \delta_{\cal C}(u_0-v_0)\delta^3(\vec u - \vec v) \,.
\label{complexgreen}
\eeq
Here $P_{L,R} = \frac{1}{2}(1 \mp \gamma^5)$ and $\theta_{\cal C}(u_0-v_0)$ and $\delta_{\cal C}(u_0-v_0)$ are the step functon and the Dirac delta function on the complex time argument. It follows in particular that the Wightmann functions 
\beqa
  iG^<(u,v) &\equiv&
         \langle  \bar \psi (v){\psi} (u)\rangle
\nonumber \\
  iG^>(u,v) &\equiv&
         \langle  {\psi}(u)\bar \psi (v)\rangle
\label{G-less-2}
\eeqa
obey the homogenous equation: 
\beq
(i\deldag_u - m^*P_L - mP_R) G^{<,>}(u,v) = 0 \,.
\label{Gless_eqn}
\eeq
Let us now solve this equation under the assumption of {\em translational invariance} both in space and in time. Translational invariance is obviously only consistent with a constant $m$, and moreover it implies:
\beq
G^{<,>}(u,v)=G^{<,>}(u-v) \,.
\label{Gtransinv}
\eeq
If we now define 
\beq
G_{\cal C}(t,{\bf x}) \equiv 
 - (i\deldag + mP_L + m^*P_R) \, \Delta_{\cal C}(t,{\bf x})
\label{GintermsofDelta}
\eeq
it is easy to see that the functions $\Delta^{<,>}$ obey the Klein-Gordon equation
\beq
(\partial_t^2 - \nabla^2 + |m|^2) \Delta^{<,>} = 0.
\label{standard}
\eeq
Making a Fourier transformation \wrt the spatial coordinate, one finds the solution
\beq
i\Delta^{<,>}(t,{\bf k}) 
= a^{<,>}_+ e^{i\omega_k t} + a^{<,>}_- e^{-i\omega_k t}\,,
\label{Dtransinv}
\eeq
where $\omega_k = \sqrt{\vec k^2+|m|^2}$. The four independent coefficient functions (of $k_0$) $a_\pm^{<,>}$ can be solved in terms of, say $a_+^>$, using the equation (\ref{Gless_eqn}) and the identity $\Delta^>(t,\vec k) = \Delta^<(-t,\vec k)$. The most general solution consistent with translational invariance is then: 
\begin{figure}
\centering
\includegraphics[width=9cm]{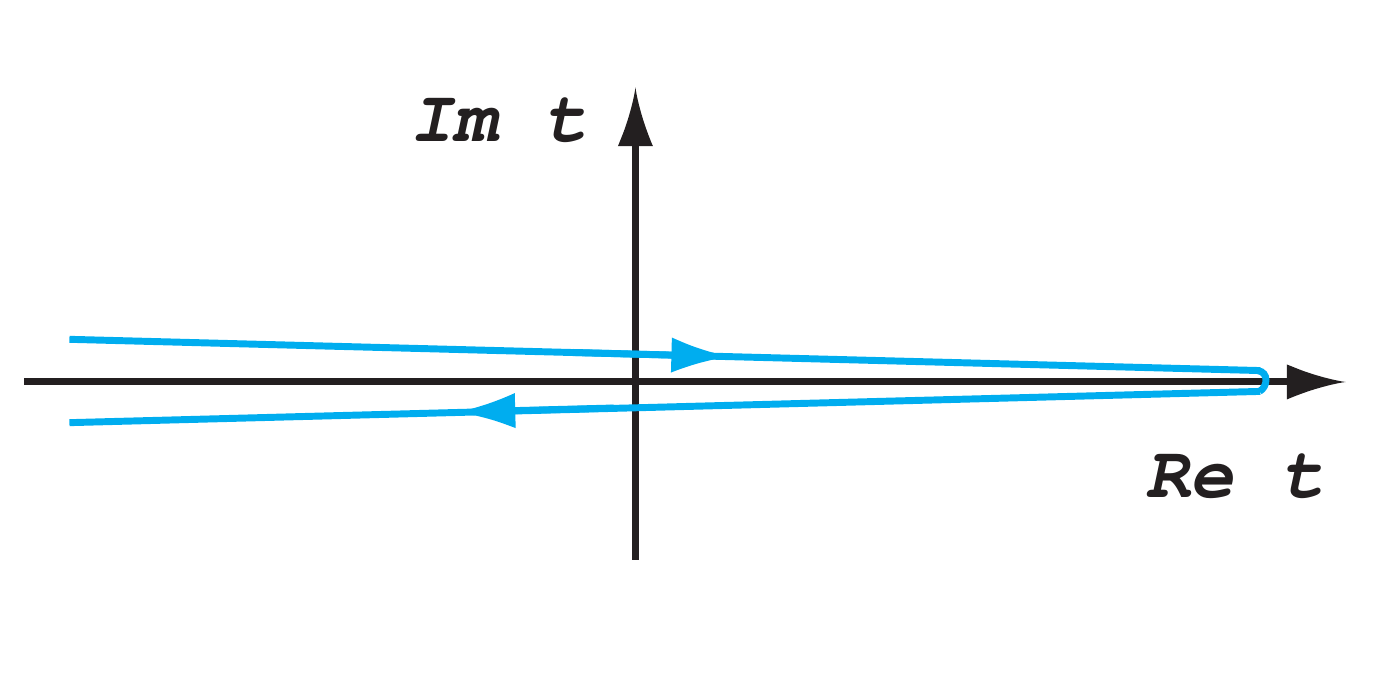}
    \vskip-1cm
    \caption{Standard complex closed time path (CTP) contour for Keldysh propagators.}
    \label{fig:keldysphath}
\end{figure}
\beq
a^>_- = a^>_+ - \frac{1}{2\omega_k} \quad {\rm and } \quad 
a^<_\pm = a^>_\mp \,,
\label{apm_solutions}
\eeq
where $a_+^>$ is the only yet unspecified function. If one assumes further that the system is in thermal equlibrium, one imposes the Kubo-Martin-Schwinger (KMS) boundary-condition:
\beq
\Delta^>_{\rm eq}(t) \equiv \Delta^<_{\rm eq}(t + i\beta) \,.
\label{KMS}
\eeq
Using Eqs.~(\ref{apm_solutions}-\ref{KMS}) one then finds: 
\beq
a^>_+ = n_{\rm eq}(\omega_k) \equiv \frac{1}{e^{\beta \omega_k}+1}
\qquad {\rm and}Ê\qquad
\Delta^>_{\rm eq} = e^{\beta \omega_k}\Delta^<_{\rm eq}\,,
\label{eqsolution1}
\eeq
where
\beq
i\Delta^<_{\rm eq}(t,{\bf k}) =  
\frac{1}{2\omega_k}n_{\rm eq}(\omega_k) ( e^{i\omega_k t} +  e^{\beta\omega_k} e^{-i\omega_k t}) \,.
\label{eqsolution2}
\eeq
Using Eqs.~(\ref{eqsolution1}-\ref{eqsolution2}) in (\ref{GintermsofDelta}) one can write down the Mills representation for the full thermal propagator:
\beq
iG_{\cal C}(t,{\bf k}) =  \int \frac{{\rm d}k_0}{2\pi }\, {\cal A }(k_0, {\bf k}) \; \big[ n_{\rm eq}(k_0) + \theta_{\cal C}(t) \big] \; e^{ik_0 t} \,,
\label{Mills}
\eeq
where we have defined the {\em spectral function}
\beq
 {\cal A} = \pi {\rm sgn}(k_0) (\kdag +  m_R - i\gamma^5 m_I)
            \delta(k^2-|m|^2) \,.
\label{specA_eka} 
\eeq
From the expression (\ref{specA_eka}) one immediately finds the standard equilibrium propagators:
\begin{equation}
 iG^<_{\rm eq} =  2 {\cal A} \, n_{\rm eq}(k_0) \,  \quad {\rm and}Ê\quad
 iG^>_{\rm eq} =  2 {\cal A} \,  (1 - n_{\rm eq}(k_0)) \,.
\label{P_gthermal}
\end{equation}
Note that $-2i{\cal A} = G^>_{\rm eq} + G^<_{\rm eq}$ as expected by the equal time anticommutator relations (see section (\ref{sec:spectral}) below). We stress again that two crucial assumptions were necessary in arriving to the equilibrium expressions (\ref{Mills}-\ref{P_gthermal}): the translational invariance both in space and in time and the standard KMS relation (\ref{KMS}).

\subsection{Free fields in varying backgrounds}

Let us now assume that the mass is some nontrivial function of space and time $m= m(t,{\bf x})$. Equation (\ref{Gless_eqn}) still holds, but we can no longer use the translational invariance to simplify the problem.  Instead, we can separate the dependence on the internal and external degrees of freedom by defining the {\em Wigner-transformation} as a Fourier transformation of a 2-point function \wrt the internal co-ordinate $r \equiv u-v$:
\beq
G(k,x) \equiv \int d^{\,4} r \, e^{ik\cdot r}
                                         G(x + r/2,x-r/2) \,,
\label{wigner1}
\eeq
where $x=(u+v)/2$ is the average co-ordinate, and $k$ is the internal momentum variable conjugate to $u-v$.  Transforming (\ref{Gless_eqn}) in this way gives:
\beq
  (\kdag + \frac{i}{2} \deldag_x
             -  \hat m_0 - i\hat m_5\gamma^5) G^{<,>}(k,x) = 0,
\label{G-lessEq2}
\eeq
where $\hat m_0$ and $\hat m_5$ are operators related to the real and imaginary parts of the mass function:
\beq
\hat m_{\rm 0,5} G^{<,>}(k,x) \equiv m_{\rm R, I}(x) e^{-\frac{i}{2}
       \partial_x^m \cdot \partial_k^G} G^{<,>}(k,x)\,.
\label{massoperators}
\eeq
Note that the derivative $\partial_x^m$ operates only to the left, acting on the mass functions $m_{\rm R,I}$ and the derivative $\partial_k^G$ acts only to the right, on function $G^{<,>}$.  Thus Eq.~(\ref{G-lessEq2}) contains an infinite number of derivative operators acting on functions $G^{<,>}$.

The dynamical functions $G^{<,>}$ and the spectral function ${\cal A}$ are not Hermitian. It is desirable to work with Hermitian functions however, and to this end we will define:\footnote{Note that refs.~\cite{KPSW,PSW} use a slightly different definition for the Hermitian function: $G \rightarrow i\gamma^0G$.  Both definitions are equally correct, although they obey slightly different equations of motion. Our present convention is convenient in that $iG^<\gamma^0$ is more directly related to the usual density matrix in Dirac indices.}
\beq
\bar G^{<,>}(u,v) \equiv i G^{<,>}(u,v)\gamma^0.
\eeq
which are easily seen to be Hermitian in the sense that:
\begin{eqnarray}
  \left[\bar G^{<,>}(u,v)\right]^\dagger &=& \bar G^{<,>}(v,u)
  \nonumber\\
  \left[\bar G^{<,>}(k,x)\right]^\dagger   &=& \bar G^{<,>}(k,x).
\label{CEq1}
\end{eqnarray}
It will also be convenient to write the equations of motion in the Weyl basis where the gamma-matrices are given by the following direct product expressions:
\beq
\gamma^0 = \rho^1 \otimes 1 \,,
\qquad
\vec \alpha = -\rho^3 \otimes \vec \sigma
\quad {\rm and} \quad \gamma^5 = -\rho^3 \otimes 1 \,.
\label{gammamatrices}
\eeq
Here both $\rho^i$ and $\sigma^i$ are the usual Pauli matrices such that the
$\rho$-matrices refer to the chiral- and $\sigma$-matrices to the spin-degrees
of freedom. In this representation, multiplying (\ref{G-lessEq2}) from both sides by $\gamma^0$, we find the equation:
\beq
\Big( 
  k_0 + \frac{i}{2}\partial_t  
+ \rho^3 \otimes \big[ \vec \sigma \cdot (\vec k - \frac{i}{2}\vec \nabla)\big]  
- ( \rho^1 \hat m_0 - \rho^2 \hat m_5)\otimes 1_2
\Big) \bar G^{<,>}(k,x) = 0 \,.
\label{G-lessEq-weyl}
\eeq
This is the master equation for this paper. It is difficult to analyse it further in full generality. However, Eq.~(\ref{G-lessEq-weyl}) can be simplified by introducing certain space-time symmetries. The {\it spatially homogenous} case may be of interest to some applications in the early universe, when the spatial gradients may be neglected, but an evolution of some background field or the expansion of the universe creates nontrivial time dependence. This is the relevant limit for example for particle production in the early universe~\cite{TomGar}, coherent baryogenesis~\cite{cohbaryog},
or for the description of the neutrino-mixing in the early universe~\cite{KKEKT}. The {\it static} (or stationary) {\it case with a planar symmetry} is relevant for the electroweak baryogenesis studies. We will consider these special cases in sections \ref{sect:tcase} and \ref{sect:zcase} below, starting from the spatially homogenous time-dependent case. 

Before going to the special cases let us make the following genaralization to our master equation. As is well known, one can always consider the fundamental chiral fermions massless and introduce the masses through interactions, technically corresponding to insertion of the singular self-energy corrections. (This is actually the best way to understand the emergence of the spatially dependent mass terms.)  However, we can also introduce interactions with other types of classical background fields by including the appropriate singular self-interaction terms. In this way we can us to generalize our master equation (\ref{G-lessEq2}) to:
\beq
  (\kdag + \frac{i}{2} \deldag_x
             - \Sigma_{\rm sing} (x) e^{-\frac{i}{2}
       \partial_x^\Sigma \cdot \partial_k^G} ) G^{<,>}(x,k) = 0.
\label{G-lessEq3}
\eeq
where $\Sigma_{\rm sing}(x)$ represents all relevant singular self-interaction corrections. We shall need this more general form when we consider the case of a quantum reflection from a potential wall in section \ref{kleinproblem}. For now however, we will mostly concentrate to the case where $\Sigma_{\rm sing} (x)$ reduces to the complex mass instertion (\ref{G-lessEq2}).

\section{Spatially homogeneous case}
\label{sect:tcase}

Let us first consider the case of a spatially homogenous system, where the translational invariance in space is restored. As a result the 3-momentum is conserved and the spatial gradient terms vanish in Eq.~(\ref{G-lessEq-weyl}), giving rise to a much simpler equation for $G^<$:
\beq
\Big( k_0 + \frac{i}{2}\partial_t
    - \vec \alpha \cdot \vec k 
    - \gamma^0 \hat m_0 - i\gamma^0 \gamma^5 \hat m_5
\Big) \bar G^<(k,t) = 0 \,,
\label{G-lessEq3hom}
\eeq
with
\beq
\hat m_{\rm 0,5} \bar G^<(k,t) \equiv m_{\rm R, I}(x) e^{-\frac{i}{2}
       \partial_t^m \partial_{k_0}^G} \bar G^<(k,t)\,.
\label{t_massoperators}
\eeq
Of course $\bar G^>$ obeys an identical equation. Homogeneity also implies that {\em helicity} is a good quantum number. This can be seen from the fact that the helicity operator $\hat h= \hat k \cdot \vec S = \hat k \cdot \gamma^0 \vec \gamma \gamma^5$, where $\hat k \equiv \vec k/|\vec k|$, commutes with the differential operator of Eq.~(\ref{G-lessEq3hom}). This fact is particularily transparent in the Weyl basis (\ref{gammamatrices}), where the helicity operator is just $\hat h = 1 \otimes \hat k \cdot \vec \sigma$. As a result one can introduce a block-diagonal decomposition for $\bar G^<$ in the helicity basis:
\beq
  \bar G_h^<  \equiv g_h^< \otimes
  \frac12(1 + h \hat k\cdot \vec \sigma),
\label{connectionHOMOG}
\eeq
where $g_h^<$ are unknown $2\times 2$ matrices in chiral indices. When the decomposition (\ref{connectionHOMOG}) is inserted into Eq.~(\ref{G-lessEq3hom}) it breaks into two independent equations (for $h = \pm 1$) for the $g_h^<$-matrices:
\beq
\Big( k_0 + \frac{i}{2}\partial_t
       + h |\vec k| \rho^3
       - {\hat m}_0 \rho^1 + {\hat m}_5 \rho^2
\Big) g_h^< = 0.
\label{Gs-eomHOMOG}
\eeq
This is as far as one can simplify the equation by using the homogeneity. 
It is still useful to rewrite Eq.~(\ref{Gs-eomHOMOG}) in a form that separates the explicit dependence on $\partial_t g^<_h$. Taking the Hermitian and anti-Hermitian parts from (\ref{Gs-eomHOMOG}) one finds two independent equations:
\begin{eqnarray}
2k_0 g^<_h &=& \hat H g_h^< + g_h^< \hat H^\dagger 
\label{Hermitian22}
\\ 
i \partial_t g^<_h &=& \hat H g_h^< - g_h^< \hat H^\dagger \, ,
\label{AntiHermitian22}
\end{eqnarray}
where  
\begin{equation}
\hat H \equiv - h|\vec k|\rho^3 + \hat m_0 \rho^1 - \hat m_5\rho^2\,. 
\label{GenHamiltonian}
\end{equation}
Note that the operators $\hat m_{0,5}$, and therefore also the operator $\hat H $ are in general not Hermitian. Equations (\ref{Hermitian22}-\ref{AntiHermitian22}) receive a particularily nice interpretation in the mean field limit (where $\hat H $ becomes Hermitian), as will be described below.

An alternative form, which we will find useful in our analysis, can be found by introducing the Bloch-representation for Hermitian $g_h^<$:
\beq
g^<_h \equiv \frac12 \left( g^h_0 +  g^h_i \rho^i \right) \,.
\label{gleh}
\eeq
In this formulation $g^<_h$ is represented by a real 4-component vector instead of a Hermitian 2x2 matrix. Using the Bloch-representation Eq.~(\ref{Gs-eomHOMOG}) can be written as:
\def\hi{\phantom{i}}
\begin{eqnarray}
(k_0 + \frac{i}{2}\partial_t) g^{h}_0  + \hi h |\vec k| g^{h}_3
            - \hi {\hat m}_0 g^{h}_1 + \hi {\hat m}_5 g^{h}_2 &=& 0
\nonumber \\
(k_0 + \frac{i}{2}\partial_t) g^{h}_3  + \hi  h |\vec k| g^{h}_0
            - i   {\hat m}_0 g^{h}_2 - i   {\hat m}_5 g^{h}_1 &=& 0
\nonumber \\
(k_0 + \frac{i}{2}\partial_t) g^{h}_1  - i    h |\vec k| g^{h}_2
            - \hi {\hat m}_0 g^{h}_0 + i   {\hat m}_5 g^{h}_3 &=& 0
\nonumber \\
(k_0 + \frac{i}{2}\partial_t) g^{h}_2  + i    h |\vec k| g^{h}_1
            + i   {\hat m}_0 g^{h}_3 + \hi {\hat m}_5 g^{h}_0 &=& 0.
\label{eq-sigmaHOMOG}
\end{eqnarray}
These equations can again be separated into two independent equations by taking the real and imaginary parts.  Formally these equations can be written as
\begin{eqnarray}
i\partial_t g^{h}_\alpha &=& \hat A_{\alpha\beta} g^h_\beta
\label{Hermitianvec}
\\
0 &=& \hat B_{\alpha\beta} g^h_\beta \,,
\label{AntiHermitianvec}
\end{eqnarray}
where the matrix operators $\hat A_{\alpha\beta}$ and $\hat B_{\alpha\beta}$
are easily read off from Eq.~(\ref{eq-sigmaHOMOG}). Equations (\ref{Hermitianvec}) and (\ref{AntiHermitianvec}), respectively, carry the same information as do the Equations (\ref{Hermitian22}) and (\ref{AntiHermitian22}).

\subsection{Mean field limit}

Transforming the original equation (\ref{Gless_eqn}) for the two point function $G^<(u,v)$ to the mixed representation given by (\ref{wigner1}) resulted in equations that contain arbitrary orders of derivative operators.  
As such it is difficult to obtain any full solutions even after using the spatial homogeneity.  To proceed further, we now consider the case where the gradient expansions are truncated to the zeroth order, {\ie} the {\em mean field limit}.  In this case the mass terms are no longer operators and the anti-Hermitian equation (\ref{Hermitian22}) becomes 
\beq
\partial_t g^<_h = -i[H, g^<_h] \,,
\label{rhoHOMOG}
\eeq
while the Hermitian one (\ref{AntiHermitian22}) reduces to
\beq
2k_0 g^<_h = \{H, g^<_h\} \,,
\label{CEHOMOG}
\eeq
where the mean field limit of the operator $\hat H$,
\beq
H = \left( \begin{array}{cc}
                   -h|\vec k| & m \\
                   m^* & h|\vec k|
                 \end{array}\right) \,,
\label{Hamilton}
\eeq
is immediately identified as the local Hamiltonian of the system. Anti-Hermitian component equations are often called {\em kinetic equations} (KE), whereas the Hermitian ones, which in the mean field limit are algebraic, are called {\em ``constraint equations''} (CE) \cite{KPSW}. Indeed, since equations (\ref{rhoHOMOG}-\ref{CEHOMOG}) constitute 8 equations for 4 scalar quantities it is natural to interpret some of the equations as constraints on the phase space in which the dynamical solution is to be found~\cite{KPSW}. 

The anti-Hermitian equation (\ref{rhoHOMOG}) looks very promising, since it clearly has just the standard form of the equation of motion that one would derive for the density matrix $\rho_h = \psi_h\psi_h^\dagger$ from the Dirac equation. However, it must be warned that interpreting it as a dynamical equation for $g^<_h$, or even interpreting $g^<_h$ as a density matrix, is not at all straightforward.  Indeed, we will next find that as a result of constraint equations (\ref{CEHOMOG}) the matrix $g^<_h$ acquires a nontrivial singular structure, whereby the Eq.~(\ref{rhoHOMOG}) is not even well defined as such.

\subsection{Shell structure, homogenous case}
\label{subsec:t-shells}
Let us first study the information contained in the constraint
equations (\ref{CEHOMOG}) in the mean field limit.  The novel result of this section will be that the constraint equations allow, in addition to the usual free particle states, a class of apparently energy conservation breaking solutions. These solutions live on the shell $k_0=0$ and we interpret them as holding the information about the quantum coherence of mixing particle and antiparticle states (zitterbewegung). The complete shell structure imposed by the constraint equations is most easily seen by first rewriting them in the form (\ref{AntiHermitianvec}), where  $B_{\alpha\beta}$ now is a constant matrix. We find
\beqa
k_0 g^h_0 + h|\vec k| g^h_3 - m_R g^h_1 + m_I g^h_2 &=& 0
\nonumber \\
k_0 g^h_3 + h|\vec k| g^h_0 &=& 0
\nonumber \\
k_0 g^h_1 - m_R g^h_0 &=& 0
\nonumber \\
k_0 g^h_2 + m_I g^h_0 &=& 0 \,.
\label{Hset1}
\eeqa
This set of equations has nontrivial solutions only when $\det (B_{\alpha\beta})=0$. It is easy to see that in the homogenous case under investigation here, this condition becomes
\begin{equation}
\det(B_{\alpha\beta}) = k_0^2(k^2-|m|^2) = 0
\end{equation}
That is, in addition to the usual mass-shell solutions with $k^2-|m|^2 =0$, there are new solutions living on $k_0=0$. Let us now find out the precise form of these solutions.

\subsubsection{$k_{0}\neq 0$ -solutions, mass-shell states}

From Eq.~(\ref{Hset1}) it is easy to see that for $k_0Ê\neq 0$ the constraint equations (\ref{Hset1}) have the following solution:
\beq
  g^h_3 = - \frac{h|\vec k|}{k_0} g^h_0, \qquad g^h_1 = \frac{m_R}{k_0} g^h_0,
  \qquad g^h_2 = - \frac{m_I}{k_0} g^h_0
\label{g13HOM}
\eeq
and
\beq
 (k_0^2 - |\vec k|^2 - |m|^2) g^h_0 = 0.
\label{SpecEqHOM}
\eeq
This equation has the spectral solution:
\beq
g^h_0(k_0,|\vec k|;t) 
   =2\pi \, f^h_{s_{k_0}}(|\vec k|,t) \,\delta(k_0 - s_{k_0}\omega_k ) \,,
\label{SpecSolHOM}
\eeq
where $s_{k_0} \equiv {\rm sgn}(k_0)$ and the mass-shell energies are given by
\beq
\omega_k \equiv \sqrt{\vec k^2 + |m|^2}.
\label{DR1HOM}
\eeq
Using (\ref{g13HOM}) and (\ref{SpecSolHOM}) we can write the full chiral mass-shell $g^<_h$-matrix as follows:
\beq
g^<_{h,{\rm m-s}}(k_0,|\vec k|;t) = 
2 \pi |k_0| \, f^h_{s_{k_0}}(|\vec k|,t) 
                                  \left(\begin{array}{cc}
                                  1 - h|\vec k|/k_0  &  m/k_0 \\
                                  m^*/k_0 &  1 + h|\vec k|/k_0
                                  \end{array} \right)
                                 \delta(k^2 - |m|^2).
\label{simpleqsk0}
\eeq
This solution has the expected form of a density matrix in the helicity eigenbasis. Exactly analogous solution exists for $g^>_{h,{\rm m-s}}$. Invoking the usual Feynman-Stueckelberg interpretation the solutions with negative ($s_{k_0}=-1$) energies can be identified with antiparticles. Finally the two unknown functions $f^h_{\pm}(|\vec k|,t)$ are generalized particle and antiparticle phase space densities. Note that in the limit $m \rightarrow 0$ we get the usual result that \eg left chirality equals negative helicity for particles and positive helicity for antiparticles.

\subsubsection{$k_{0}=0$ -solutions, zitterbewegung}

The mass-shell solutions (\ref{simpleqsk0}) were derived assuming that $k_0 \neq 0$. However, setting $k_0=0$ in the first place, but keeping $|\vec k| \neq 0$, we find out that equations (\ref{Hset1}) have a new class of solutions, which obey the relations
\begin{eqnarray}
g_3^h & = & h\frac{m_R}{|\vec k|} g_1^h - h\frac{m_I}{|\vec k|} g_2^h
\nonumber \\
g_0^h & = & 0,
\label{q-cohHOMOG}
\end{eqnarray}
while the components $\bar g_{1,2}^h$ are unconstrained. The corresponding
spectral solution is
\begin{eqnarray}
 g^<_{h,{\rm 0-s}}(k_0,|\vec k|;t) &=&  \pi \left[
                   f^h_{1}(|\vec k|,t) \left( \begin{array}{cc}
                                  h\, m_R/|\vec k| &  1 \\
                                  1       & - h\, m_R/|\vec k|
                                 \end{array} \right)
                             \right.
  \nonumber \\    && \phantom{m} + \left.
                   f^h_{2}(|\vec k|,t) \left( \begin{array}{cc}
                                  -h\, m_I/|\vec k| &  -i \\
                                  i       &  h\, m_I/|\vec k|
                                 \end{array} \right)
                    \right] \,
                    \delta (k_0) \,,
\label{k0zerospec}
\end{eqnarray}
where $f^h_{1}(|\vec k|,t)$ and $f^h_{2}(|\vec k|,t)$ are two unknown real functions living on the shell $k_0 = 0$, and so they cannot be directly associated with either particles or antiparticles. However, since one expects that a general time-dependent density matrix should contain information of the quantum coherence between particles and antiparticles, we make the following identification: {\em the additional $k_0=0$-solutions (\ref{k0zerospec}) describe the quantum coherence (zitterbewegung) between particles and antiparticles with same helicity $h$ and opposite momenta $\vec k$}.

\begin{figure}
\centering
\hskip -3truecm
\includegraphics[width=11cm]{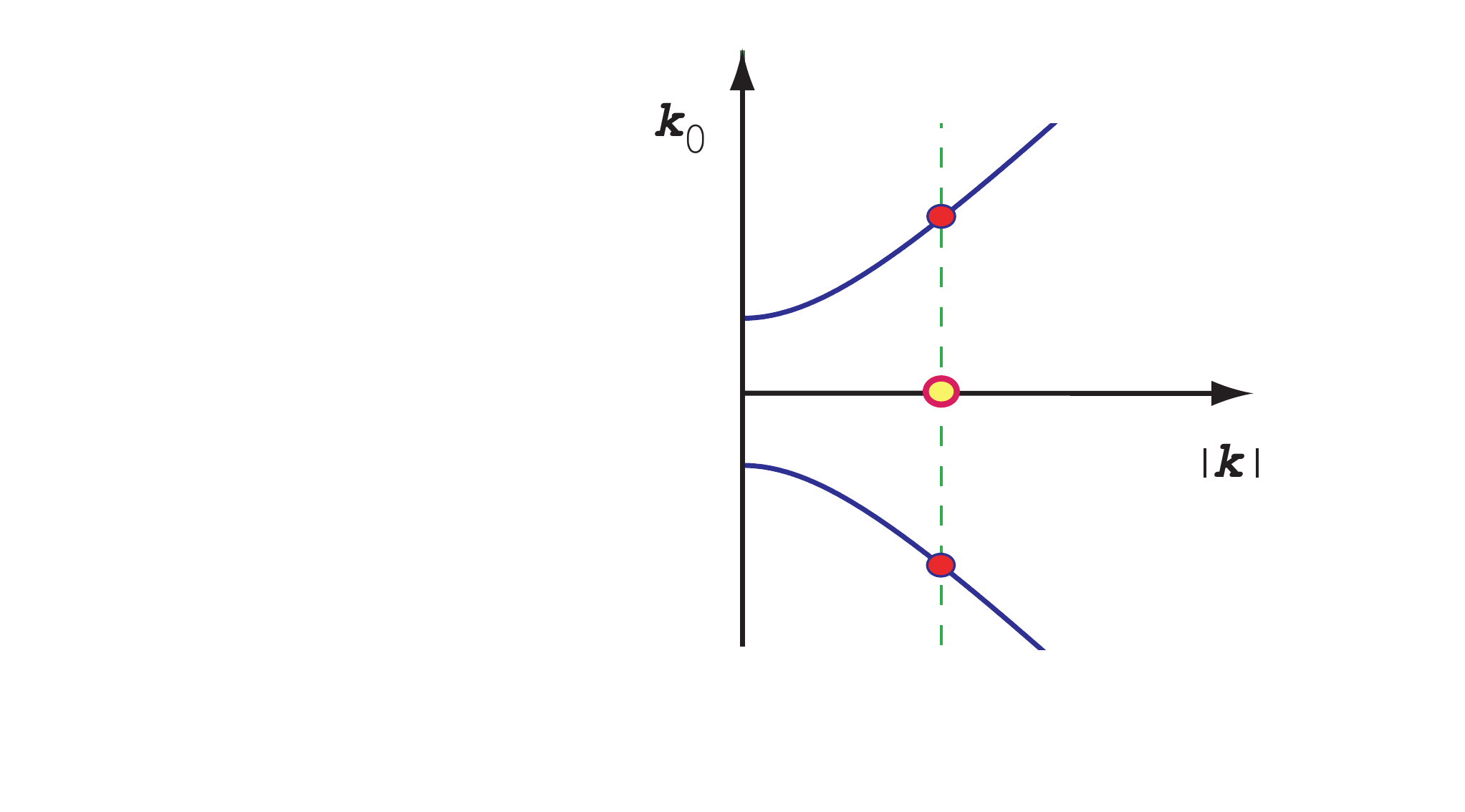}
\vskip -1.2cm 
    \caption{Dispersion relations in the case of a scalar mass function. 
    The dark filled (red) blobs show the mass-shell contributions of a given $|\vec k|$ to the
    matrix $g^<_h$ and the light (yellow) blob shows the corresponding coherence 	
    solution from the new $k_0=0$-shell.}
    \label{fig:DR-omega}
\end{figure}

\noindent
The most complete solution for a given momentum $|\vec k|$ and helicity $h$ can be written as
\beq
  g^<_h(k_0, |\vec k|;t) = g^<_{h,{\rm m-s}}(k_0,|\vec k|;t) \, + \, g^<_{h,{\rm 0-s}}(k_0,|\vec k|;t) \,.	
\label{fullgh}
\eeq
The full solution (\ref{fullgh}) contains {\em four} independent spectral functions $f^h_{\alpha}(|\vec k|,t)$ living on three distinct shells. These shells are represented in the phase space diagram in figure \ref{fig:DR-omega}. It is this singular structure which appears to render the evolution equation (\ref{rhoHOMOG}) to be of little use. This ambiguity is only lifted when one interprets $g_h^<$ as a phase space density, and defines the true physical density matrix as a weighted integral over the singular $g_h^<$. We shall postpone introducing these ideas more precisely until section \ref{sect:dynamicaleqns}. For now let us show that our new $k_0=0$-solutions are not present in the spectral function and in the usual thermal limits for $G^{<,>}$ that follow by use of the KMS-relations. 

\subsection{Spectral function}
\label{sec:spectral}

The spectral function is defined in general as
\beqa
  {\cal A}(u,v) \equiv
        \frac{1}{2} \langle  \{ \psi (u), \bar \psi(v) \} \rangle 
      = \frac{i}{2}(G^>(u,v) + G^<(u,v)) \,.
\label{G-less}
\eeqa
In the free field case ${\cal A}(u,v)$ obeys the same equation of motion as functions $G^{<,>}$, and so the most general solution for ${\cal A}$ is of the form of Eq.~(\ref{connectionHOMOG}):
\beq 
  {\cal A} \gamma^0 = \sum_h a_h \otimes \frac12(1 + h \hat k\cdot
  \vec \sigma) \,, 
\label{specA1HOMOG} 
\eeq
where the chiral matrix $a_h$ is identical to the most general solution (\ref{fullgh}) for $g^<_h$, with four yet undefined spectral on-shell functions $f^{h \cal A}_\alpha$ for both helicities. However, in the homogenous case we are implicitly assuming that there are no nontrivial spatial correlations, so that we can perform the usual quantization by imposing the equal time anticommutation rules for the field operators:
\beq
\{\psi(t,\vec u), \psi^\dagger (t,\vec v)\} = -i \delta^3(\vec u-\vec v) \,.
\label{equalcomm}
\eeq
It is easy to see that these anticommutation relations imply that
\beq
2 {\cal A}(t,\vec u; t, \vec v) \gamma^0 =  \delta^3(\vec u-\vec v)\,.
\label{spectrel1}
\eeq
This condition is just the direct space version of the famous {\em spectral sum-rule}, whose mixed representation counterpart reads as:
\beq
  \int \frac{{\rm d}k_0}{\pi}
                  {\cal A}(k,x) \gamma^0  = 1.
\label{sumruleHOM}
\eeq
It turns out that the sum-rule (\ref{sumruleHOM}) is enough to completely fix the values of the on-shell
functions $f^{h \cal A}_\alpha$:
\beq 
f^{h \cal A}_\pm = \frac12 \quad\quad {\textrm{and}} \quad\quad f^{h \cal A}_{1,2} = 0 \,, 
\label{specAconstrHOMOG}
\eeq
for both helicities. With these values the full solution for ${\cal A}$ becomes:
\beq
 {\cal A} = \pi {\rm sgn}(k_0) (\kdag +  m_R - i\gamma^5 m_I)
            \delta(k^2-|m|^2) \,.
\label{specA3} \eeq
This is just the familiar result for the spectral function in thermal quasiparticle limit found in section \ref{sect:standardCTPprop}, see also for example~\cite{PSW}. The spectral function is thus completely determined, and it does not contain any dynamics at all.  Moreover, it does not have any contribution from the $k_0=0$-shell describing the coherence between particles and antiparticles.  This is what one should expect: since coherence is a {\em dynamic} phenomenon, it should not show up in the measure of the one-particle phase space. This is precisely what we are seeing here. Moreover, it should be vanishing in the statistical equilibrium limit.

\subsection{Equilibrium limit for $G^{<,>}$}

The a priori independent distributions $f^{<,>}_{s_{k_0},h}$ in functions $G^<$ and $G^>$ are constrained by the relation $G^> + G^< = - 2i {\cal A}$. Using Eq.~(\ref{specAconstrHOMOG}) we then find:
\beqa
f^{h <}_{s_{k_0}} + f^{h >}_{s_{k_0}} &=& 1 \nonumber \\
f^{h <}_{(1,2)} + f^{h >}_{(1,2)} &=& 0 \,.
\label{relationshere}
\eeqa
(Note that we drop the $<,>$-indices on $f^h_\alpha$-functions everywhere where there is no danger of confusion.) Relations (\ref{relationshere}) hold generally, as long as the spectral solutions are valid. However, if one further imposes thermal equilibrium by the Kubo-Martin-Schwinger (KMS) boundary condition 
\beq
G^>_{\rm eq}(t) \equiv G^<_{\rm eq}(t + i\beta)  \qquad \Rightarrow \qquad 
G^>_{\rm eq}(k_0) = e^{\beta k_0}G^<_{\rm eq}(k_0) \,,
\eeq
one can easily show that the quantum coherence functions must vanish:
\beq
f^{h <,>}_{(1,2)} = 0
\label{eqf12}
\eeq
and the mass-shell distributions become:
\beqa
f^{h <}_{s_{k_0}} &=& n_{\rm eq}(k_0) \nonumber \\
f^{h >}_{s_{k_0}} &=& 1- n_{\rm eq}(k_0)  \,,
\label{eqfspm}
\eeqa
where $n_{\rm eq}(k_0) = 1/(e^{\beta k_0} +1)$ is the usual Fermi-Dirac distribution.  It is now easy to write down the equilibrium solutions for $G^{<,>}$: 
\beq
  \bar G^{<,>}_{\rm eq}  \equiv \sum_h g_{{\rm eq},h}^{<,>} \otimes
  \frac12(1 + h \hat k\cdot \vec \sigma) \,.
\label{fullGsolution}
\eeq
Using solutions (\ref{eqf12}-\ref{eqfspm}) in (\ref{simpleqsk0}), (\ref{k0zerospec})  and (\ref{fullgh}) equation (\ref{fullGsolution}) reduces to:
\begin{eqnarray}
 iG^<_{\rm eq} &=& 
       2\pi {\rm sgn}(k_0) (\kdag +  m_R - i\gamma^5 m_I) \,n_{\rm eq}(k_0) \,
            \delta(k^2-|m|^2) 
\nonumber \\
 iG^>_{\rm eq} &=&  2\pi {\rm sgn}(k_0) (\kdag +  m_R - i\gamma^5 m_I)
         \,  (1 - n_{\rm eq}(k_0)) \, \delta(k^2-|m|^2) \,,
\label{SK_gthermal}
\end{eqnarray}
which are just the standard equilibrium propagators found in section \ref{sect:standardCTPprop}.

The crucial difference between our treatment and the standard derivation of the thermal propagator is how we treat the space-time symmetries. In the usual approach the coherence-solutions are excluded already before imposing the KMS-relations by the assumption of translational invariance. We can easily see how this limit arises in our approach. Eqs.~(\ref{Hermitian22}-\ref{AntiHermitian22}) are already translationally invariant in space by homogeneity. Imposing also the time-translational invariance $\partial_t g^h_\alpha \equiv 0$ (and $\partial_t m \equiv 0$) turns the kinetic Eqs.~(\ref{AntiHermitian22}) into four additional algebraic constraints. These constraints are consistent with the mass-shell solutions (\ref{simpleqsk0}) as long as $f^h_\pm$ are constants. Coherence solutions are inconsistent with them however, and hence the coherence is directly excluded by translational invariance.  In our more complete treatment the functions $f^h_{s_{k_0}}(|\vec k|,t)$ can be  time-dependent and the quantum coherence is maintained in the form of the dynamical functions $f^h_{(1,2)}(|\vec k|,t)$. 

\section{Planar symmetric case}
\label{sect:zcase}

One often encounters situations where quantum states interact with classical backgrounds that can be approximated by planar configurations. Examples range from simple quantum reflection problems to particle interactions with an expanding phase transition wall during electroweak baryogenesis. Let us now assume that the system is symmetric along planes orthogonal to the $z$-axis. In this case the equation of motion (\ref{G-lessEq-weyl}) becomes
\beq
\Big( \,  k_0 + \frac{i}{2}\partial_t
-  \alpha^3 \, (k_z - \frac{i}{2}\partial_z)  - \vec \alpha  \cdot \vec k_\pp
        - \gamma^0  \hat m_0  - i\gamma^0\gamma^5 \hat m_5
        \, \Big) \, \bar G^<(k;t,z) = 0 \,.
\label{G-lessEq4}
\eeq
Unlike in the homogenous case, helicity is not conserved here. However, one
notices that apart from $\vec \alpha_{\pp } \cdot \vec k_\pp$-term the
differential operator in Eq.~(\ref{G-lessEq4}) commutes with the spin in
$z$-direction, which is described by the operator $S^3 =
\gamma^0\gamma^3\gamma^5 = 1 \otimes \sigma^3$. One can try to get rid of the $\vec \alpha_\pp$-terms by boosting to a frame where all reference to $\vec k_\pp$ vanishes. Putting aside the transformation of the derivative-operators, the boost $\Lambda_\pp$ should obviously be such that
\beq
 S(\Lambda_\pp)  \kdag  S^{-1}(\Lambda_\pp)
   \equiv \tilde k_0 \gamma^0 - k_z \gamma^3 \,,
\label{boostframe}
\eeq
where $\tilde k_0 = {\rm sgn(k_0)}({k_0^2-\vec k_\|^{\,2}})^{\frac 12}$. The explicit form of the boost matrix $S(\Lambda_\pp)$ is easy to work out:
\beq
  S(\Lambda_\pp) = {\rm sgn(k_0)} \frac{k_0 + \tilde{k}_0
               - \vec{\alpha}\cdot\vec{k}_\pp}
                {\sqrt{2\tilde{k}_0(k_0+\tilde{k}_0)}}.
\label{boostperp}
\eeq
The boost $\Lambda_{\pp }$ obviously leaves the form of the derivative operator invariant: $\deldag = \gamma^0\partial_{t'} - \vec \gamma_{\pp }\cdot \partial_{x'_\pp } -  \gamma^3\partial_{z}$. However, as the boost mixes the time and space components, the planar symmetry argument
\beq
\partial_{\vec x_\pp}  \bar G^<(k;t,z) = 0
\label{planarsymm}
\eeq
does not hold anymore in the new coordinates. Instead, one can show that in the new coordinates
\beq
\vec \alpha \cdot \partial_{\vec x_\pp'}  \bar G'^<(k;x') =
\gamma_\pp \, \vec \alpha \cdot \vec v_\pp \; \partial _t \bar G'^<(k;t,z),
\eeq
where $\vec v_\pp \equiv \vec k_\pp/k_0$ and $\gamma_\pp \equiv 1/(1-v_\pp^{2})^{1/2}.$  That is, the boost regenerates the noncommuting $\vec \alpha \cdot \vec k_\pp$-terms from the gradients and the boosted differential operator still fails to commute with $S^3$ in general. An obvious exception to this rule is the {\em static case} where also $\partial _t \bar G \equiv 0$.  In the static case the boost (\ref{boostperp}) {\em does} remove all dependence on $\vec \alpha_\pp$ from equation (\ref{G-lessEq4}), and reduces the differential operator block-diagonal in the spin along the $z$-axis~\cite{KPSW}. In what follows, we will restrict the analysis to the static, or more generally {\em stationary} cases. The latter can always be reduced to a static problem by a suitable Lorentz tranformation, as we shall see next.

\subsection{Stationary problems with planar symmetry}

In the application to the electroweak baryogenesis one can assume that the planar symmetric background fields have a {\em stationary} dependence on $t$ and $z$. In particular for the mass function one can assume a form
\beq
m(t,z) = m_w(z - v_w t) \,,
\eeq
where $v_w$ is the velocity of the phase transition front in the {\it plasma frame}.  This stationary form implies that the mass function is static in the {\it wall frame}, which is connected to the plasma frame by a Lorenz-transform
\begin{eqnarray}
  t_w = \gamma_w (t - v_w z ) &\qquad& k_{0_w} = \gamma_w (k_0 - v_w k_z )
\phantom{ \, ,}
\nonumber \\ 
  z_w = \gamma_w (z - v_w t ) &\qquad& k_{0_z} = \gamma_w (k_z - v_w k_w ) \, ,
\end{eqnarray}
where $\gamma_w \equiv 1/({1-v_w^2})^{1/2}$. (That is: $m(t,z) = m_w(z_w/\gamma_w) \equiv m(z_w)$.) This boost is a constant in momentum variables, and hence leaves the mass-operators invariant. The spinor representation of the transform $\Lambda_w$ is
\beq
  S(\Lambda_w) = \frac{1}{\sqrt{2}}
         \big(\sqrt{\gamma_w+1}
             -\sqrt{\gamma_w-1} \; \alpha^3
         \big).
\label{boostvw}
\eeq
The boost $S(\Lambda_w)$ obviously commutes with $\vec \alpha_\pp$. In the wall frame the boosted function
\beq
\bar G^<_w(k; z_w) \equiv
 S(\Lambda_w)\, \bar G^<(k;z-v_wt) \,
 S(\Lambda_w)
\eeq
obeys a static equation
\beq
\Big( \,  k_{0_w}
- \alpha^3 \, (k_{z_w} - \frac{i}{2}\partial_{z_w})
- \vec \alpha  \cdot \vec k_\pp
        - \gamma^0  \hat m_{0}  - i\gamma^0\gamma^5 \hat m_{5}
        \, \Big) \, \bar G_w^<(k; z_w) = 0 \,.
\label{G-lessEq5_static}
\eeq
where in particular the mass operators are static in the wall frame variables:
\beq
\hat m_{0,5} = m_{\rm R, I}(z_w)
e^{\frac{i}{2}{\partial}^m_{z_w}\partial^G_{k_{z_w}}} \, .
\eeq
Because the problem is static in the wall frame, Eq.~(\ref{G-lessEq5_static}), it can be boosted to a frame where all $\vec \alphaÊ\cdot \vec k_\pp$-terms vanish. The explicit spinor representation $S(\Lambda_{\pp w})$ of the boost is given by Eq.~(\ref{boostperp}) with $k_0 \rightarrow k_{0_w}$.  After this second boost one finds that the function
\beq
  \bar G^<_{\pp w}(\tilde k_{0_w}, k_{z_w};z_w)
    \equiv S(\Lambda_{\pp w})  \bar G_w^<(k;z_w) S(\Lambda_{\pp w})
\eeq
obeys the equation,
\beq
\Big( \,  \tilde k_{0_w}
-  \alpha^3 \, (k_{z_w} - \frac{i}{2}\partial_{z_w})
        - \gamma^0  \hat m_{0}  - i\gamma^0\gamma^5 \hat m_{5}
        \, \Big) \, \bar G_{\pp w}^<(\tilde k_{0_w}, k_{z_w}; z_w) = 0 \,,
\label{G-lessEq5_noperb}
\eeq
where $\tilde k_{0_w} = {\rm sgn}(k_0) (k_{0_w}^2-k_\pp^2)^{1/2}$. In the doubly boosted frame, one can make use of the commutativity of the differential operator with $S^3$ and introduce the spin-decomposition analogous to Eq.~(\ref{connectionHOMOG}):
\beq
\bar G_{\pp w,s}^<  \equiv g_{\pp w,s}^< \otimes \frac12(1 + s \sigma_z) \, ,
\label{connection}
\eeq
where $g_{\pp w,s}^<$ are two (for $s= \pm 1$) unknown Hermitian $2 \times 2$ matrices in chiral indices. Inserting (\ref{connection}) into Eq.~(\ref{G-lessEq5_noperb}) one finds
\beq
\Big( {\tilde k}_{0_w}
       + s (k_{z_w} - \frac{i}{2}\partial_{z_w}) \rho^3
       - {\hat m}_0 \rho^1 + {\hat m}_5 \rho^2
\Big) g_{\pp w,s}^< = 0  \, .
\label{Gs-eom}
\eeq
This equation is very similar to Eq.~(\ref{Gs-eomHOMOG}). If one identifies $sk_{z_w}$ with $h|\vec k|$ (and $\tilde k_{0_w}$ with $k_0$), the only difference is replacing $\partial_t$-operator by an operator $-s\rho^3\partial_z$. This small change leads to profoundly different solutions however.

\subsection{Division to constraint and {\em evolution} equations}
\label{divisiontoevoeq}

We now proceed to analyse (\ref{Gs-eom}) in a same manner as we analysed Eq.~(\ref{Gs-eomHOMOG}) in section \ref{sect:tcase}. However, a direct division of Eq.~(\ref{Gs-eom}) into Hermitian and anti-Hermitian equations does {\em not} lead to the desired separation to kinetic and constraint equations~\footnote{Note that in the earlier work related to electroweak baryogenesis~\cite{KPSW} the conceptually wrong division to constraint and kinetic equations -- following Hermiticity properties -- was used. While certainly wrong for the discussion of the quantum reflection, we have checked that this choice in the end does not affect the semiclassical limit discussed in these papers.}. Instead, one has to first multiply the equation from left by $\rho^3$ and only {\em then} perform the division. In this way we find the equations
\begin{eqnarray}
-2 s k_z g^<_s &=& \hat P g_s^< + g_s^< \hat P^\dagger
\label{Hermitian22z}
\\ 
is \partial_z g^<_s &=& \hat P g_s^< - g_s^< \hat P^\dagger \, ,
\label{AntiHermitian22z}
\end{eqnarray}
where
\begin{equation}
\hat P \equiv k_0 \rho^3  + i(\hat m_0 \rho^2 + \hat m_5\rho^1)\,. 
\label{Pamilton}
\end{equation}
Here we have dropped all indices referring to wall frame or to the frame with zero parallel momentum. One can either assume that boosts have been done, or that we consider the case with either $v_w=0$ or $\vec k_\pp=0$, or both. Obviously, the operator $\hat P$ is a generalization of a local momentum operator in the same manner as $\hat H$ generalized the local Hamiltonian. Alternatively, introducing again the Bloch representation
\beq
g^<_s \equiv \frac12 \left( g^s_0 +  g^s_i \rho^i \right) \,,
\label{glesss}
\eeq
we can decompose equation (\ref{Gs-eom}) into components as follows:
\begin{eqnarray}
 s(k_z - \frac{i}{2}\partial_z) g^s_0 + \hi k_0 g^s_3 
           - i {\hat m}_0 g^s_2 - i {\hat m}_5 g^s_1 &=& 0
\nonumber \\
 s(k_z - \frac{i}{2}\partial_z) g^s_3 + \hi k_0 g^s_0
           - \hi {\hat m}_0 g^s_1 + \hi {\hat m}_5 g^s_2 &=& 0
\nonumber \\
 s(k_z - \frac{i}{2}\partial_z) g^s_1  - i k_0 g^s_2  
            + \hi  {\hat m}_0 g^s_3 - i {\hat m}_5 g^s_0 &=& 0
\nonumber \\
 s(k_z - \frac{i}{2}\partial_z) g^s_2 + i k_0 g^s_1
            - i {\hat m}_0 g^s_0  - \hi {\hat m}_5 g^s_3 &=& 0 \,.
\label{eq-sigma-static}
\end{eqnarray}
Again, these equations can be divided to real and imaginary parts, resulting in vector equations $is\partial_z g^s_\alpha = \hat A^z_{\alpha\beta}g^s_\beta $ and $0 = \hat B^z_{\alpha\beta}g^s_\beta$, where $\hat A^z_{\alpha\beta}$ and 
$\hat B^z_{\alpha\beta}$ are operators that can easily be read off from Eq.~(\ref{eq-sigma-static}) and which become simple constant matrices in the mean field limit.

Unlike $\hat H$ the operator $\hat P$ is not Hermitian even in the mean field limit. It is easy to see that it nevertheless takes the familiar form of the local momentum operator
\beq
P   \equiv \left( \begin{array}{cc}
                   k_0 & -m \\
                   m^* & -k_0
                 \end{array}\right),
\label{Peq}
\eeq
while the constraint and evolution equations become:
\begin{eqnarray}
-2 s k_z g^<_s &=& P g_s^< + g_s^< P^\dagger
\label{Hermitian22zb}
\\ 
is \partial_z g^<_s &=& P g_s^< - g_s^<P^\dagger \, .
\label{AntiHermitian22zb}
\end{eqnarray}
One again recognizes that the evolution equation has the standard form of the equation for a density matrix $\rho = \psi\psi^\dagger$, which can be simply derived using the static Dirac equation~\cite{CKV}.  

\subsection{Shell structure, planar case}
\label{subsec:z-shells}
The analysis of the shell structure proceeds very similarly to the homogenous case.  The new result will be that the constraint equations allow a class of apparently {\em momentum} conservation breaking solutions in addition to the usual free particle states. These solutions live on the shell $k_z=0$ and turn out to hold the information about the quantum coherence between mixing opposite momentum states with same spin. The shell structure becomes evident by solving the determinant condition for the mean field limit constraint equations:
\beqa
k_0 g^s_0 + sk_z g^s_3 - m_R g^s_1 + m_I g^s_2 &=& 0
\nonumber \\
k_0 g^s_3 + sk_z g^s_0 &=& 0
\nonumber \\
sk_z g^s_1 + m_R g^s_3 &=& 0
\nonumber \\
sk_z g^s_2 - m_I g^s_3 &=& 0 \, .
\label{Cset1}
\eeqa
It is easy to see that the determinant of this set of equations is just
\begin{equation}
\det(B^z_{\alpha\beta}) = k_z^2 (k^2 - |m|^2) \, .
\end{equation}
So, by setting the determinant to zero we find again similar on-shell solutions as we did in the homogenous case, but instead of $k_0=0$ solution, we now find a double root at $k_z=0$. Let us now find the explicit $g^<_s$ matrices corresponding to these solutions.

\subsubsection{$k_{z}\neq 0$ -solutions; free particle-shells}

It is again easy to show that for $k_z \neq 0$ equations (\ref{Cset1})
have the particlular solution
\beq
  g^s_3 = - \frac{sk_z}{k_0} g^s_0, \qquad g^s_1 = \frac{m_R}{k_0} g^s_0,
  \qquad g_2 = - \frac{m_I}{k_0} g^s_0 \, ,
\label{g13}
\eeq
and
\beq
 (k_0^2 - k_z^2 - |m|^2) g^s_0 = 0 \, .
\label{SpecEq}
\eeq
Equation (\ref{SpecEq}) has the spectral solution
\beq
  g^s_0(k_0,k_z;z)  = 
             2\pi \, f^s_{s_{k_z}}(k_0,z) \,\frac{|k_0|}{|k_z|} \,\delta(k_z - s_{k_z} k_m ) \, ,
\label{SpecSol}
\eeq
where $s_{k_z} \equiv {\rm sgn}(k_z)$. These solutions thus live on a well defined energy-momentum shell corresponding to the usual dispersion relation
\beq
  k_m =  \sqrt{k_0^2 - |m|^2}.
\label{DR1}
\eeq
Note that we are taking energy as the free variable, whereas $k_z$ is defined by the on-shell condition. This is the appropriate choice for a problem with spatial gradients, where the momentum need not be conserved. Using (\ref{g13}) and (\ref{SpecSol}) we can write corresponding full chiral $g^<$-matrix as follows:
\beq
  g^<_{s,\rm m-s}(k_0,k_z;z) = 2 \pi |k_0| \, f^s_{s_{k_z}}(k_0,z) 
                            \left( \begin{array}{cc}
                              1 - sk_z/k_0  &  m/k_0 \\
                              m^*/k_0       &  1 + sk_z/k_0
                            \end{array} \right)
                             \delta(k^2 - |m|^2).
\label{simpleqs}
\eeq
This solution again has the expected form of an mass-shell state of a definite spin in $z$-direction, and again, an analogous solution exists for $g^>_{s,{\rm m-s}}$. The unknown functions $f^s_{\pm}(k_0,z)$ will be identified as generalized phase space densities for right and left moving states.

\subsubsection{$k_{z}=0$-solutions and quantum coherence}

The free particle solutions (\ref{simpleqs}) were derived assuming that $k_z \neq 0$. Giving up this restriction, we find that equations (\ref{Cset1}) allow a new class of solutions living on shell $k_z=0$. Still keeping $k_0 \neq 0$, we find the solution:
\begin{eqnarray}
g_0^s & = & \frac{m_R}{k_0} g_1^s - \frac{m_I}{k_0} g_2^s
\nonumber \\
g_3^s & = & 0,
\label{q-coh}
\end{eqnarray}
while the components $g_{1,2}^s$ are unconstrained.  The corresponding
spectral solution is
\begin{eqnarray}
 g^<_{s,{\rm 0-s}}(k_0,k_z;z) &=&  \pi \left[
                      f^s_{1}(k_0,z) \left( \begin{array}{cc}
                                  m_R/k_0 &  1 \\
                                  1       &  m_R/k_0
                                 \end{array} \right)
                             \right.
  \nonumber \\    && \phantom{m} + \left.
                   f^s_{2}(k_0,z) \left( \begin{array}{cc}
                                  -m_I/k_0 &  -i \\
                                  i       &  -m_I/k_0
                                 \end{array} \right)
                    \right] \,
                    \delta (k_z) \,,
\label{kzzerospec}
\end{eqnarray}
where $f^s_{1}$ and $f^s_{2}$ are unknown functions that only depend on the energy and the position.  To see what physics these new solutions describe,  note first that each state of a definite energy $k_0$ and spin $s$ can correspond to two different states with opposite helicities and momenta. Secondly, a density matrix should carry information about the quantum coherence between degenerate states, that may be present when the defining quantum numbers (here the momentum) are sufficiently poorly known. No combination of free particle solutions (\ref{simpleqs}) can carry such information however. We thus make the following interpretation: \textit {the additional $k_z$=0-shell solutions (\ref{q-coh}) describe the quantum coherence of states of equal spin travelling in opposite directions}. This is also very natural from the momentum conservation point of view: while the mixing mass shell components have large and opposite momenta $k_z = \pm \sqrt{k_0^2-m^2}$, their coherent mixture has the momentum expectation value of $k_z=0$. 

\begin{figure}
\centering
\includegraphics[width=8cm]{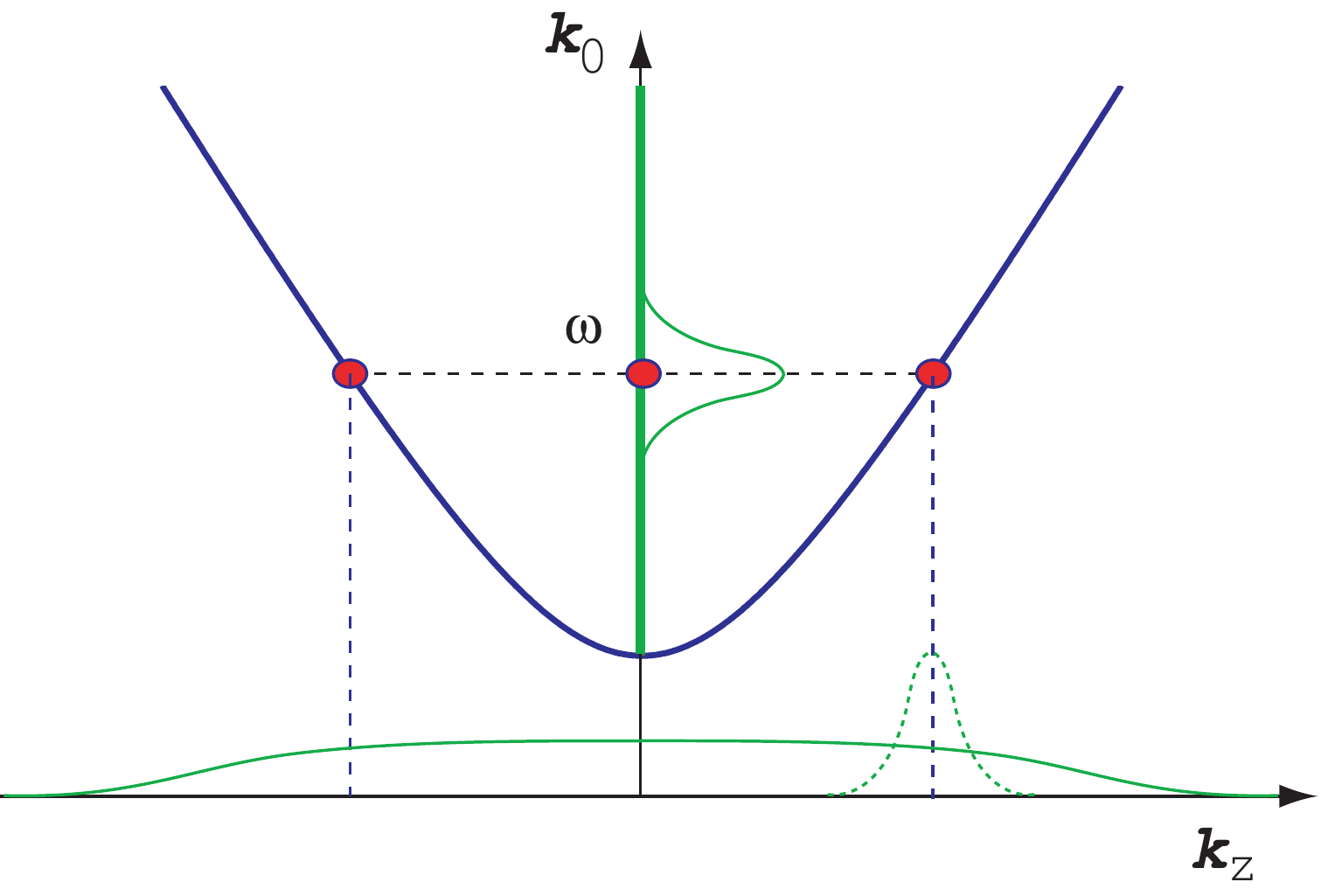}
    \caption{Dispersion relation in the case of a scalar mass term. The
    filled blobs show the on-shell contributions to the matrix $g^<_s$.
    Peaked functions on each axis illustrate the accuracy of the exterior
    knowledge on $k_0$ and $k_z$. In the case of $k_z$ the dashed line
    function illustrates an effectively accurate measurement of momentum
    direction and the solid line function the case of a complete
    ignorance on the direction.}
    \label{fig:ScalarDR}
\end{figure}

\noindent The most complete solution satisfying the constraint equations (\ref{Cset1}) for a given spin $s$ and energy $k_0 \neq 0$  is
\beq
  g^<_s(k_0,k_z;z) = g^<_{s,{\rm m-s}}(k_0,k_z;z) \, + \, g^<_{s,{\rm 0-s}}(k_0,k_z;z) \,.
\label{fullchiral}
\eeq
The practical use of this solution is again restricted by the fact that the  matrices describing the physical mass-shell solutions and their quantum coherence are proportional to distributions defined on different momentum shells in the phase space. These dispersion relations with particluar solutions for a given $k_0$ are shown in figure \ref{fig:ScalarDR}. We will discuss the physical interpretation of these spectral solutions in section \ref{sect:dynamicaleqns}. Before entering this discussion we wish to end this section by showing how the coherence solutions change if one considers a stationary instead of a static problem: that is if one considers the problem in the plasma- rather than in the wall frame.

\subsection{Finite $v_w$; stationary case in wall frame}

Since the purpose of this section is merely to illustrate the effect of
stationarity on the $k_z=0$-solution, we shall restrict ourselves to the
1+1-dimensional case here (\ie we set $\vec k_\pp= 0$). In that case the relevant equations of motion
can be read from (\ref{G-lessEq4}) with constant mass terms and replacement $\partial_t \rightarrow - v_w \partial_z$. Following our division to the constraint and kinetic equations in the static case we now find
\beqa
k_0 g^s_0 + sk_z g^s_3 -  m_R g^s_1 + m_I g^s_2 &=& 0
\nonumber \\
k_0 g^s_3 + sk_z g^s_0 &=& 0
\nonumber \\
-v_w\partial_z g^s_1 - 2sk_z g^s_2 +  2m_I g^s_3 &=& 0
\nonumber \\
-v_w\partial_z g^s_2 + 2sk_z g^s_1 + 2m_R g^s_3 &=& 0 \,,
\label{statioCons}
\eeqa
and
\beqa
s \partial_z g^s_3 &=& - v_w\partial_z  g^s_0 
\nonumber \\
s \partial_z g^s_0 &=& - 2 m_R g^s_2 - 2  m_I g^s_1 - v_w\partial_z  g^s_3 
\nonumber \\
s \partial_z g^s_1 &=& -2k_0 g^s_2 - 2 m_I g^s_0  
\nonumber \\
s \partial_z g^s_2 &=& \phantom{-}
                        2k_0 g^s_1 - 2 m_R g^s_0 \,.
\label{statioKin}
\eeqa
At first sight the situation appears problematic, since we now have only
two purely algebraic constraint equations left in (\ref{statioCons}). 
However, since we have four independent linear equations for the two 
derivatives $\partial_z g^s_{1,2}$, we can use the (last two) kinetic 
equations (\ref{statioKin}) to eliminate these derivatives from constraints. 
When this is done, the last two equations in (\ref{statioCons}) become:
\beqa
s(k_z-v_wk_0) g^s_1 + m_R ( g^s_3 + s v_wg^s_0) &=& 0
\nonumber \\
s(k_z-v_wk_0) g^s_2 - m_I ( g^s_3 + s v_wg^s_0)  &=& 0 \,.
\label{statioCons2}
\eeqa
We thus are back to four independent algebraic constraint equations.
Using the second constraint in (\ref{statioCons}), which still gives (when $k_0 \neq 0$) 
$g^s_3 = -\frac{sk_z}{k_0}g^s_0$, the equations (\ref{statioCons2})
can be rewritten as
\beqa
  s (k_z-v_wk_0) \left(g^s_1 - \frac{m_R}{k_0} g^s_0 \right) &=& 0
\nonumber \\
  s (k_z-v_wk_0) \left(g^s_2 + \frac{m_I}{k_0} g^s_0 \right) &=& 0 \, .
\label{statioCons3}
\eeqa
The presence of the new coherence shell solution is in fact more apparent
here than it was in the static case: from (\ref{statioCons3}) one  
immediately sees that if $k_z-v_wk_0 \neq 0$ then 
$g^s_1 = \frac{m_R}{k_0} g^s_0$ and $g^s_2 = - \frac{m_I}{k_0} g^s_0$,
which immediately leads to the usual mass-shell solutions.
However, in addition the {\em combinations} 
$\tilde g^s_1 \equiv g^s_1 - \frac{m_R}{k_0} g^s_0$ and 
$\tilde g^s_2 \equiv g^s_2 + \frac{m_I}{k_0} g^s_0$ have spectral 
solutions $\tilde g^s_{1,2} \propto \tilde f^s_{1,2}\delta(k_z-v_w k_0)$. 
While the new shell structure is most obviously visible in variables 
$\tilde g^s_{1,2}$ it is a simple matter of linear mapping to replace
the free on-shell functions $\tilde f^s_{1,2}$ with those associated with 
the distributions $g^s_{1,2}$.  The coherence shell has thus moved to
$k_z = w_wk_0$ as it should by the Lorentz transformation between the
static and stationary frames.  Setting $k_z=v_wk_0$ in the first two
constraint equations then gives the $g^s_{0,3}$ in terms of the free
variables $g^s_{1,2}$ on the coherence shell:
\beq
  g^s_3 = -sv_wg_0  \quad {\rm and} \quad 
  g^s_0 = -\gamma_w^2(\frac{m_R}{k_0}g^s_1 + \frac{m_I}{k_0}g^s_2)\,.
\eeq
Finally, a direct calculation shows that the physical mass shells remain 
to be given by the Lorentz-invariant relation $k_0^2 - k_z^2 - |m|^2 = 0$.

\section{Dynamical equations and connection to the measurement theory}
\label{sect:dynamicaleqns}

In previous sections we have discovered and interpreted the complete shell structure of the free fermion propagator when a complete translational invariance is lifted either in the temporal, or in one of the spatial directions. We found in particular that in the mean field limit the propagator matrix has spectral solutions including the usual mass shell, but moreover also new solutions on shells where $k_0 = 0$ (homogenous) or $k_z=0$ (static, planar symmetry). We interpreted these shells as carrying information about the quantum coherence  between particles and antiparticles of equal helicity and opposite momenta in the homogenous case, and between left and right moving states of equal spin under reflection from a wall in the planar symmetric case. However, these interpretations are still problematic in that it is not clear what we mean by coherence, since it is living on a different singular shell. Indeed, both our final dynamical equations (\ref{rhoHOMOG}) and (\ref{AntiHermitian22zb}) are yet ill defined because of this issue. We now show how this situation is to be interpreted, and in process discover an intersting connection to the measurement theory.

\subsection{From $g^<$ to a weighted density matrix}
\label{sect:wfunc}

The key idea is that in reality we can never have a complete information on the variables describing a certain process. Hence the physically interesting object -- whose evolution we {\em can} study -- is not the singular density matrix $g^<_{s,h}$ but some smeared-out object\footnote{This description serves the purpose for the present argument, and it would be the approach to be taken in many experimental situations. In reality, the role of the measurement is taken by the collision term that couples the system to the surroundings. This is actually the approach we shall take in the later publications where we will extend the present analysis to the case with collisions~\cite{HKR,HKR2}.}, whose definition involves the extrenous information about the parameters of the system into the theory. To see how this works consider first the planar symmetric case, where we assume that we have a fairly precise information of the momentum $k_z$ as well as of the energy $k_0$ and the spin $s$ of the state at all $z$. This situation is illustrated in the figure \ref{fig:ScalarDR} for the phase space variables. This information excludes the coherence solutions and the full density matrix (\ref{fullchiral}) is reduced to the form given by equation (\ref{simpleqs}). When this structure, through relations (\ref{g13}), is fed into the evolution equation (\ref{AntiHermitian22zb}) one finds that $P g_{s,{\rm m-s}}^< - g_{s,{\rm m-s}}^<P^\dagger = 0$.
This implies that the functions $f^s_{s_{k_z}}(k_0,z)$ are constants so that the solution $g^<_s(k_0,k_z)$ describes a free particle propagation without any quantum coherence.  

The above example may look trivial, but the important issue to note is 
that a precise information, or an ideal \textit{measurement} of energy and momentum variables reduce the evolution equation (\ref{AntiHermitian22zb}) to a trivial description of a free particle propagation.  So how does the usual density matrix picture with nontrivial quantum coherence emerge?  The answer   is that the coherence is possible only when the extrenous information about the state of the system is sufficiently inaccurate. Suppose now that our prior knowledge for example on the energy, momentum and spin variables can be described by some weight function ${\cal W} (k_0,k_z,s \mid\hskip-0.5truemm\mid  k'_0,k'_z,s'\,;\,z)$, where the primed variables are free and those without primes denote their known mean values. For example ${\cal W}$ could be a normal distribution in $k_0$ and $k_z$, with variances $\sigma_0$ and $\sigma_z$:
\beq
{\cal W} = N 
e^{(k_0-k'_0)^2/2\sigma_0^2} e^{(k_z-k'_z)^2/2\sigma_z^2} \
 \delta_{s,s'} \,,
\eeq
where we still took spin to be fixed and $N$ is some normalization factor. (Other quantum numbers could of course be treated equally.) We now postulate that a physically observable density matrix can be defined in terms of the singular $g_s^<$ and the experiment related weight function as follows:
\beq
  \rho_{\cal W} (k_0,k_z,s;z)
     \equiv \sum_{s'} \int \frac{{\rm d}k'_0}{2\pi} \frac{{\rm d}k'_z}{2\pi}\;
      {\cal W} (k_0,k_z,s \mid\hskip-0.5truemm\mid k'_0,k'_z,s'\,;\,z)
      \; g^<_{s'}(k_z',k_0';z) \,.
\label{rhoW}
\eeq
First note that our first example is easily described in this language, where we implicitly used a weight function which imposes strict ideal measurements of energy, momentum and spin of the  state. As a second example, let us now assume that we have a complete ignorance on the direction of the momentum of the state, while we do have a precise information of the spin and the energy. Assuming that $k_0 \equiv \omega > 0$ this setting corresponds to a weight function
\beq
{\cal W} = 2\pi \delta(\omega- k_0' ) \, \delta_{s,s'}.
\label{weight2}
\eeq
It is now easy to see that the corresponding smeared out density matrix
\beq
\rho_{\cal W}(\omega,k_z,s;z) = \sum_{s'} \int 
\frac{{\rm d}k'_0}{2\pi} \frac{{\rm d}k'_z}{2\pi}\; 2\pi
\delta(\omega-k_0' ) \, \delta_{s,s'} g^<_{s'}(k_z',k_0';z)  
\; \equiv \;  \rho_s(\omega;z)
\label{smearedrhoex}
\eeq
obeys the standard evolution equation:
\beq
is\partial_z \rho_s = P \, \rho_s - \rho_s \, P^\dagger\,,
\label{rho2z}
\eeq
where 
\beq
P \equiv \left( \begin{array}{cc}
                 \omega & -m \\
                 m^* & -\omega
                \end{array}\right).
\label{Peq2}
\eeq
is just the operator given in Eq.~(\ref{Peq}) with $k_0$ set to the externally imposed value $k_0 \equiv \omega$.  This equation is exact, and the singular structure plaguing the Eq.~(\ref{AntiHermitian22zb}) has now been removed by integration so that Eq.~(\ref{rho2z}) indeed is just the ``normal" density matrix equation, capable of carrying information about coherent evolution. We stress again that this nontrivial structure emerged as a result of convoluting the (here rather the lack of the) available external information about the system on the definition of the physical density matrix.

Because of the singular form of $g^<_s$, the integration in (\ref{smearedrhoex}) is trivial and the components of $\rho_s$ can be directly related to the on-shell functions $f^s_\alpha$ appearing in Eqs.~(\ref{simpleqs}) and (\ref{kzzerospec}):
\begin{eqnarray}
\rho^s_{LL} &=&  \frac{1}{2}(\frac{\omega}{k_m}+s)f^s_{-}
             + \frac{1}{2}(\frac{\omega}{k_m}-s)f^s_{+}
             + \frac{m_R}{2\omega } f^s_{1} - \frac{m_I}{2\omega } f^s_{2}
\nonumber \\
\rho^s_{RR} &=&  \frac{1}{2}(\frac{\omega}{k_m}-s)f^s_{-}
             + \frac{1}{2}(\frac{\omega}{k_m}+s)f^s_{+}
             + \frac{m_R}{2\omega } f^s_{1} - \frac{m_I}{2\omega } f^s_{2}
\nonumber \\
\rho^s_{LR} &=&  \frac{m}{2k_m }(f^s_{-} + f^s_{+})
             + \frac{1}{2}(f^s_{1} - if^s_{2})
\nonumber \\
\rho^s_{RL} &=&  \frac{m^*}{2k_m }(f^s_{-} + f^s_{+})
             + \frac{1}{2}(f^s_{1} + if^s_{2}) \,,
\label{rhocomp}
\end{eqnarray}
where $f^s_{\pm}$ refer to functions $f^s_{s_{k_z}}(\omega; z)$ with $s_{k_z} = \pm 1$ in Eq.~(\ref{SpecSol}) and $f^s_{1,2}$ to $f^s_{1,2}(\omega; z)$ in Eq.~(\ref{kzzerospec}). Note that all components of $\rho^s$ mix terms from the mass- and coherence shells. One can extract the information about the particle numbers and coherence from $\rho_s$ at any point of the calculation by inverting the equations (\ref{rhocomp}).  Note that the four degrees of freedom encompassed by the combined mass-shell and coherence shells matches that of the most general Hermitian 2x2 density matrix. Without coherence shells $\rho_s$ would contain only two degrees of freedom which is insufficient to describe any nontrivial quantum mixing.

We conclude this subsection with comments related to the choice of the weight functions.  First, it should be kept in mind that relations (\ref{rhocomp}) between $\rho_{ij}$ and $f_\alpha$ are specific to the particular weight
function Eq.~(\ref{weight2}). In principle, the weight connection could be
something completely different, possibly encoding much more complicated structures of extrenous information about the system. This information could be spatially dependent (as is the case in the example in section \ref{kleinproblem}), or involve some partial, yet incomplete, information about a given quantum state. In any case, for any weight function there would always exist an in principle calculable relation connecting the two sets of variables.

Finally, we point out that a weight function similar to (\ref{weight2}) is actually the appropriate one to use for example for the interactions with a mass wall, encountered in the Electroweak baryogenesis problem. The reasoning is that in such case the {\em only} information one has about the system comes in the form of a set of conserved quantum numbers; in this case the energy, the momentum along the wall and the spin perpendicular to the wall. The density matrix can always be taken to be diagonal in conserved quantum numbers, but we can impose no extrenous constraint on a variable like $k_z$ for example.  Instead, one needs to introduce an explicit collision term which will give rise to damping terms that destroy the coherence generated by the interaction with the wall. This is of course the ultimate goal of our formalism, the results of which will be presented elsewhere~\cite{HKR}.

To illustrate the use of the physical density matrices, and the corresponding choices of the appropriate weight functions, we next use our formalism to solve two known reflection problems. These examples will also further underline the neccessity of retaining the coherence solutions in the density matrix.

\subsection{Klein problem}
\label{kleinproblem}

As our first example of the use of evolution equations (\ref{rho2z}), we shall
consider a fermion reflecting off a step potential. This is of course a well known Klein problem, whose solution is known in the Dirac equation approach. For this problem we need to use the extended version of our master equation (\ref{G-lessEq3}), with $\Sigma_{\rm sing}(u) = \gamma^0 V(u)$ where $V(u)$ is the potential appearing in the usual Dirac equation. In addition we now take the mass to be a real constant so that equation (\ref{G-lessEq3}) becomes:
\beq
(i \deldag_u - m - \gamma_0 V(u)) G^<(u,v) = 0 \, .
\label{DspaceKleineq}
\eeq
In the mixed representation one readily finds
\beq
   \big(\kdag + \frac{i}{2} \deldag_x  -  m
    - \gamma_0 V(x) e^{-\frac{i}{2}{\partial}^V_x \cdot \partial^G_k}
   \big) G^<(k,x) = 0.
\label{MspaceKleineq}
\eeq
Given this equation, we shall proceed with the analysis as in section (\ref{sect:zcase}). In the case of a step-potential the spatial gradients acting on $V$ vanish everywhere except exactly at the potential wall, and within the wall the potential can be absorbed to the energy $k_0$. Apart form the singular wall front the solution must then be of the form (\ref{fullchiral}), where $k_0 \rightarrow k_0 - V$ within the wall region. Moreover, interaction with a wall conserves the spin in $z$-direction.  Thus, in the region I, shown in figure \ref{fig:KleinWall}, the density matrix is a quantum mixture of incoming and outgoing (say) positive spin states with energy $k_0 = \omega$ and in region II it describes a single outgoing $s=1$ state with an effective energy $k_0= \omega-V$. This information can be expressed in terms of a single $z$-dependent weight fuction ${\cal W}$ as follows:\footnote{In reality energy is conserved, but with this trick we get to account for the potential in a single weight function.}
\beq
{\cal W} = 2 \pi \left[ \theta( z) \delta(k_0-\omega )
                + \theta(-z) \delta(k_0-\omega+V)
           \right]
 \delta_{s,1}.
\label{weight3}
\eeq
The explicit form for the density matrix can now be derived from Eq.~(\ref{rhoW}):
\begin{figure}
\centering
\vskip-1.5truecm
\includegraphics[width=16cm]{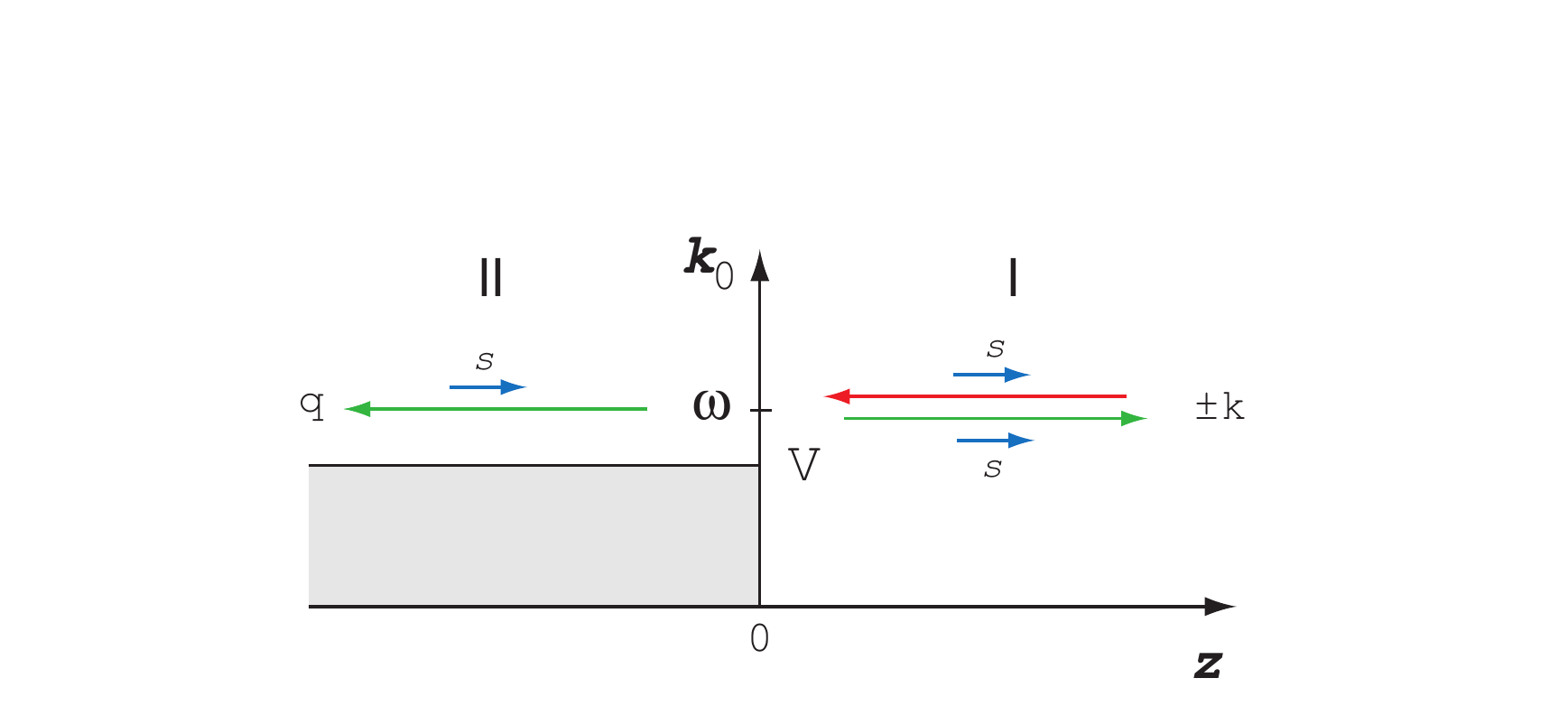}
    \caption{Reflection from a step-like potential. Long arrows describe
     the directions of momenta of the in- and outgoing particles, and the
     short vector their direction of spin.}
    \label{fig:KleinWall}
\end{figure}
\beq
\rho(z) = \theta(z) \, \rho_{\rm I} + \theta(-z) \, \rho_{\rm II} \,,
\eeq
where $\rho_{I,II}$ can be directly (leaving out an overall half in all matrices) read off from Eq.~(\ref{rhocomp}).  We find
\beq
\rho_{\rm I} =
         f^{\rm I}_-
               \left( \begin{array}{cc}
                       \frac{\omega}{k} + 1 & \frac{m}{k }  \\
                       \frac{m}{k } & \frac{\omega}{k} - 1  \\
                      \end{array}
               \right)
       + f^{\rm I}_+
               \left( \begin{array}{cc}
                       \frac{\omega}{k}-1 & \frac{m}{k }  \\
                       \frac{m}{k} & \frac{\omega}{k}+1  \\
                      \end{array}
               \right)
       + f^{\rm I}_1
               \left( \begin{array}{cc}
                       \frac{m}{\omega } & 1  \\
                        1 & \frac{m}{\omega }  \\
                      \end{array}
               \right)
       + f^{\rm I}_2
               \left( \begin{array}{cc}
                       0 & -i  \\
                       i & 0  \\
                      \end{array}
               \right)
\label{KIrho}
\eeq
where $k \equiv \sqrt{\omega^2-m^2}$. For $\rho_{\rm II}$ one replaces $\omega \rightarrow  \omega-V$ and $k \rightarrow q \equiv \sqrt{(\omega-V)^2-m^2}$. As usual there are two distinct possibilities depending on whether $q$ is real or not. For a real $q$ we expect a transmitted wave in the region II, but no incoming wave from the left, so that asymptotically $f^{\rm II}_+ = 0$.  The function $f^{\rm I}_+(z)$ corresponds to the flux of reflected states and $f^{\rm II}_-(z)$ to the flux of transmitted states, while functions $f^{\rm I,II}_{1,2}(z)$ give the coherence. Using Eqs.~(\ref{rhocomp}) we can write Eq.~(\ref{rho2z}) dircectly in terms of the $f$-functions:
\begin{eqnarray}
s\partial_z f_\pm &=& 0
\nonumber \\
s\partial_z f_1 &=& -2k_0 f_2 
\nonumber \\
s\partial_z f_2 &=& \frac{2k_z^2}{k_0}f_1 \, ,
\label{kivaeqn}
\end{eqnarray}
where $k_0 = \omega$ in the region I and $k_0 = \omega-V$ region II, and $k_z = \sqrt{k_0^2-m^2}$. We see that the functions $f^{\rm I,II}_{\pm}$ are constants, and so we can normalize the incoming flux to one in the region I: $f^{\rm I}_- \equiv 1$ and we find that $f^{\rm II}_+ = 0$ throughout. Moreover, the functions $f^{\rm I,II}_{1,2}$ are oscillatory for a real momentum. Since our boundary condition excludes asymptotic right moving states in the region II, also coherence functions must vanish there: $f^{\rm II}_{1,2} = 0$. In region I coherence is possible and one finds
\begin{equation}
f^{\rm I}_{1,2}(z) = A_{1,2} \cos(2kz) + B_{1,2}\sin (2kz) \,.
\end{equation}
The coefficients $A_2$ and $B_2$ are related to $A_1$ and $B_1$ through equations (\ref{kivaeqn}) and the remaining coefficients $f^{\rm I}_+$, $f^{\rm II}_-$, $A_1$ and $B_1$ are set by matching $\rho_{\rm I}$ and $\rho_{\rm II}$ at $z=0$. One finds that
\beq
f^{\rm I}_+ = \frac{1-x}{1+x}, \qquad {\rm where} \qquad
x \equiv \frac{qk}{\omega (\omega - V) - m^2} \,,
\label{KleinSolution1}
\eeq
the flux is conserved:
\begin{equation}
f^{\rm I}_+  + f^{\rm II}_- = 1
\end{equation}
and finally the coherence solution is:
\begin{equation}
f^{\rm I}_1(z) =  \frac{m\omega V}{k^2q}\frac{2x}{1+x} \cos(2kz) 
\qquad {\rm and} \qquad 
f^{\rm I}_2(z) = -\frac{1}{2\omega}\partial_z f^{\rm I}_1(z) \,. 
\label{KleinSolution2}
\end{equation}
This is just the familiar result known from a Dirac equation approach~\cite{ItZu}, where $f^{\rm I}_+$ and $f^{\rm II}_{-}$ are identified with the usual reflection and transmission constants. In particular when $V\rightarrow 0$ we get $q \rightarrow k$ and $x\rightarrow 1$, so that $f^{\rm I}_{+} \rightarrow 0$, $f^{\rm II}_{-} \rightarrow 1$  and $f^{\rm I}_{1,2} \rightarrow 0$ as expected.

In case when $q$ is imaginary, we cannot have any mass-shell solutions in the region II, so that $f^{\rm II}_\pm = 0$. However, we can allow coherence solutions to be nonzero there, as long as they become asymptotically zero as $z\rightarrow -\infty$. It is evident from Eq.~(\ref{kivaeqn}) that when $q$ is imaginary, appropriate exponentially decaying coherence solutions do exist. After a short calculation one finds the result with a complete reflection: $f_+^{\rm I} = 1$ and with
\beq
f^{\rm I}_1(z) = \frac{k(\omega-V)}{mV}\cos(2kz) + \frac{|q|\omega}{mV}\sin(2kz)
\eeq
and $f^{\rm I}_2(z) = -\partial_z f^{\rm I}_1(z)/{2\omega}$. In the region II one has $f^{\rm II}_\pm=0$ and the coherence functions are:
\begin{eqnarray}
f^{\rm II}_1(z) &=& \frac{k(\omega-V)}{mV} e^{2|q|z}
\end{eqnarray}
and $f^{\rm I}_2(z) = -\partial_z f^{\rm I}_1(z)/{2(\omega-V)}$. The pure coherence in this case can be interpreted as describing a virtual pair of left moving state and right moving antistate (an anti-left mover).

The lesson to be learned from this excercise is the necessity of including the $k_z=0$-shell solutions in the mixture of the states; should we have dropped them, there would have been no consistent solution to the problem at all. This is not surprising, beause leaving out $f_{1,2}$ would physically correspond to making precise measurements of the momentum content of the state at all positions, arbitrarily close to the wall.  Such measurements would significantly disturb and alter the quantum system under study.

\subsection{Reflection from a CP-violating mass wall}
\label{sect:refl}

\begin{figure}
\centering
\vskip-1.5truecm
\includegraphics[width=16cm]{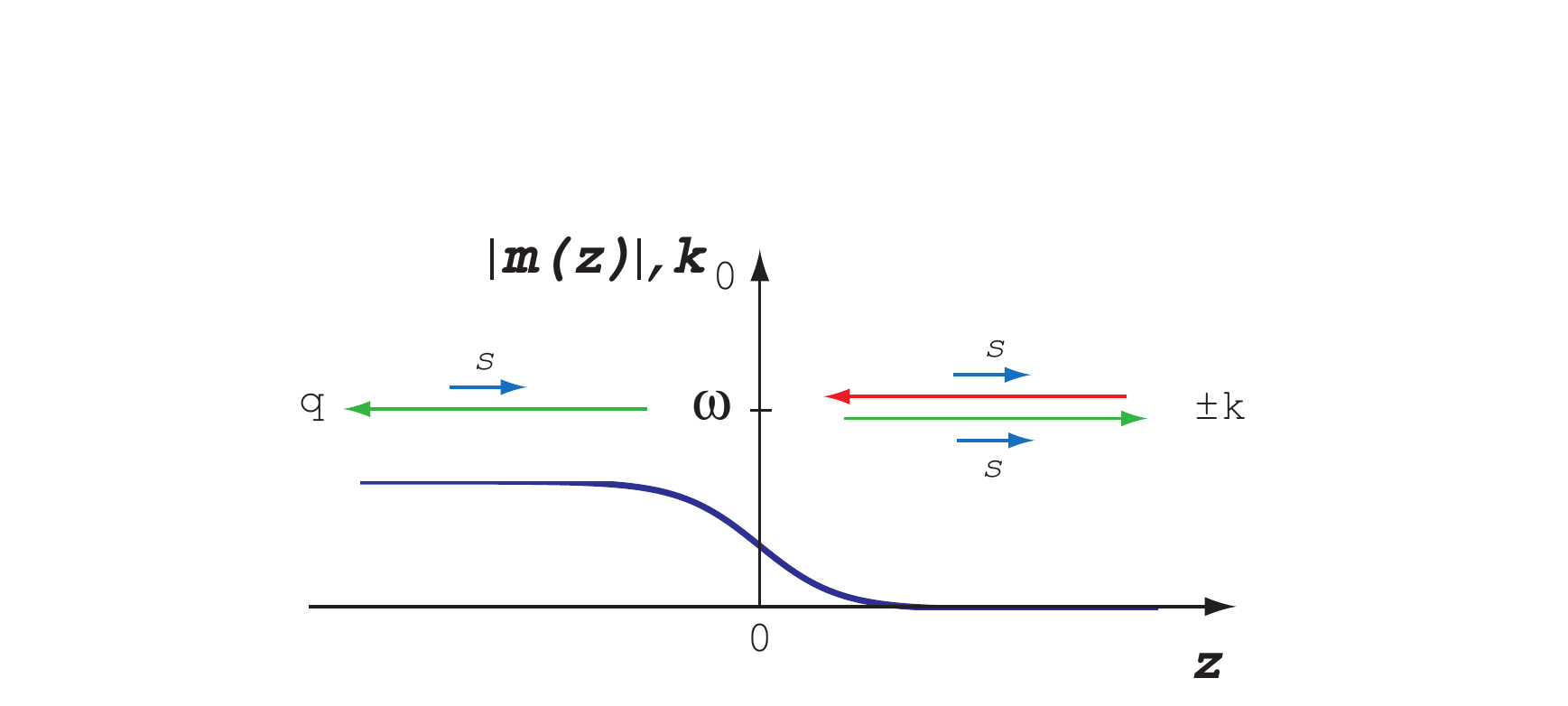}
    \caption{Reflection from a potential wall due to spatially varying
             complex mass function. Note that in this example the mass
             is not related to energy in the same way as the potential
             $V$ in the previous figure.}
    \label{fig:CPwall}
\end{figure}

As another reflection problem, we shall use our density matrix formalism
to re-derive the CP-violating chiral flux from a wall created by a spatially
varying complex mass function. This is the simplest example of a reflection
problem relevant for electroweak baryogenesis, and it has been studied in the
Dirac equation approach for example in references~\cite{FarSha,CKV}.
The setup for the problem is depicted in figure \ref{fig:CPwall}.
In this case the mass function is assumed to arise from the Higgs-mechanism,
such that
\beq
m(z) = y \phi(z)
\label{massaprof}
\eeq
where $y$ is a Yukawa coupling and $\phi(z)$ is some complex scalar
field~\cite{CKV}. Deep in the {\em unbroken phase} the field vanishes,
$\phi(z) \rightarrow 0$ as $z \rightarrow \infty$, and all particles are
massless. Far in the {\em broken phase} on the other hand, particles have a
finite mass $y\phi(-\infty) \equiv m_{-\infty}$, whose phase can be chosen
real. In the vicinity of the phase transition wall mass function is complex
and spatially varying. To be specific, we have used the following
parametrization for $\phi(z)$:
\beq
|\phi(z)| = \frac{1}{2}(1-{\rm tanh}(z/\ell_w))
\qquad
\arg \phi(z) = \frac{1}{2}\Delta\theta (1+{\rm tanh}(z/\ell_w)) \, ,
\label{phiprof}
\eeq
where $\ell_w$ is the width of the wall, and $\Delta \theta$ is the total magnitude of the change of the phase of  $\phi$.
This problem can be described by equations (\ref{Peq}-\ref{Cset1}). The density matrix appropriate for the problem is set by the weight function (\ref{weight2}), because all we know is that energy and spin are conserved quantities. Thus the equation of motion for our density matrix in chiral basis is just Eq.~(\ref{rho2z}) where the density matrix components are given by Eq.~(\ref{rhocomp}). In practice it is more convenient to employ the Bloch-representation for the physical density matrix:
\begin{equation}
\rho_s \equiv \frac{1}{2} \big( \langle g^s_0 \rangle
+ {\langle \vec g^s \rangle} \cdot \vec \sigma \big) \,, \qquad
\langle g^s_\alpha \rangle \equiv \int \frac{{\rm d}k_z}{2\pi} g^s_\alpha(k_z,\omega;z) \,.
\end{equation}
In this representation the equation of motion (\ref{rho2z}) becomes
\beqa
s \partial_z \langle g^s_0 \rangle &=& - 2 m_R \langle g^s_2 \rangle - 2  m_I \langle g^s_1\rangle  
\nonumber \\
s \partial_z \langle g^s_1\rangle &=& -2k_0 \langle g^s_2\rangle - 2 m_I \langle g^s_0\rangle  
\nonumber \\
s \partial_z \langle g^s_2\rangle &=& \phantom{-}
                 2k_0 \langle g^s_1 \rangle - 2 m_R \langle g^s_0 \rangle 
\nonumber \\
s \partial_z \langle g^s_3 \rangle &=& \phantom{-} 0 \, ,
\label{statioKinave}
\eeqa
and the variables $\langle g^s_\alpha \rangle$ are related to on-shell functions
$f^s_{\alpha}$ ($\alpha=\pm$, 1 or 2) as follows:
\begin{eqnarray}
\langle g^s_0 \rangle &=& \frac{\omega}{k_m}(f^s_{-} + f^s_{+}) 
            +\frac{m_R}{\omega}f^s_1  - \frac{m_I}{\omega}f^s_{2}
\nonumber \\
\langle g^s_1 \rangle &=& \frac{m_R}{k_m}(f^s_{-} + f^s_{+}) + f^s_{1}
\nonumber \\
\langle g^s_2 \rangle &=& -\frac{m_I}{k_m}(f^s_{-} + f^s_{+}) + f^s_{2}
\nonumber \\
\langle g^s_3 \rangle &=&  s(f^s_{-} - f^s_{+}) \,.
\label{gsintermsoff}
\end{eqnarray}
It is easy to solve equations (\ref{statioKinave}) with a simple shooting algorithm. We take the initial conditions to be such that the incoming particle flux is normalized to unity, and that no flux is coming from the left. In terms of the on-shell functions $f^s_{\alpha}$ these conditions correspond to:\footnote{Indeed, note that with the normalization (\ref{simpleqs}), we find that the fermionic current is related to $f^s_{\pm}$:
\begin{eqnarray}
\langle j^3 \rangle
 = \langle \bar\psi \gamma^3 \psiÊ\rangle 
 &=& \int \frac{{\rm d}^4k}{(2\pi )^4} {\rm Tr}[\alpha^3 \bar G^< ] 
 = -\sum_s s \int \frac{{\rm d}k_0{\rm d}^2k_{\pp}}{(2\pi )^3} 
                           \langle g^s_3 \rangle
\nonumber \\
 &=&  \sum_s \int \frac{{\rm d}k_0{\rm d}^2k_{\pp}}{(2\pi )^3} 
                           (f^s_{+} - f^s_{-})Ê\,,
\end{eqnarray}
That is, $f^s_{\pm}$ are to be interpreted as {\em flux densities per unit energy and perpendicular momentum volume} in the phase space. This is the appropriate interpretation for a problem where $k_z$ is not conserved globally. However, since on mass-shell and in the mean field limit ${\rm d}k_0 = v_z {\rm d}k_z$, the functions $f^s_{\pm}$'s can also be understood as local particle number densities per local unit 3-momentum.}
\begin{figure}
\centering
\includegraphics[width=14cm]{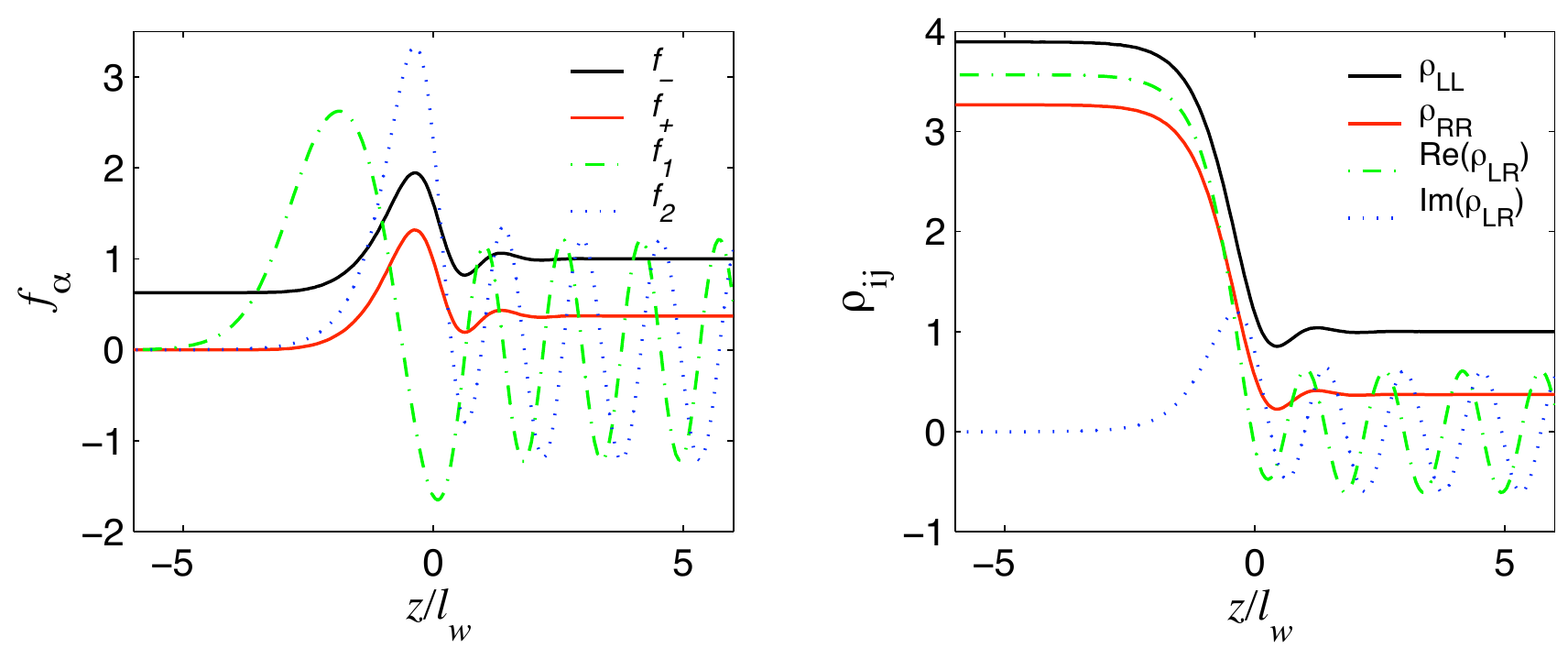}
    \caption{a) Shown are the mass-shell functions $f_\pm$ corresponding to the
             left and right moving fluxes and functions $f_{1,2}$, which 
             encode the quantum coherence. We have taken $s=1$ and 	 	
             $q/|m_{-\infty}|=0.088$.
             b) The chiral density matrix components for
             the same solution. The wall parameterization is the same as in
             figure \ref{fig:deltaj}.}
    \label{fig:fprofs}
\end{figure}
\begin{eqnarray}
f_- =  1, &\quad & z \rightarrow \infty
\nonumber \\
f_+ = f_1 = f_2 = 0, &\quad & z \rightarrow -\infty \,.
\label{bc}
\end{eqnarray}
where we have set $s=1$.
In figure \ref{fig:fprofs}a we plot the flux-functions $f_\pm$ along with the coherence functions $f_{1,2}$ for the case where the asymptotic momentum to mass ratio is $q/|m_{-\infty}| = 0.088$. From Fig.~(\ref{fig:fprofs}a) one can see that to the right from the wall the system is a coherent superposition of left and right moving states with opposite momenta, and the $k_z$=0-shell functions are oscillating coherently. In the broken phase however, all but the $f_-$-function die off and the state soon becomes a pure transmitted left moving state. Note that the physical flux-normalization condition $f_+(\infty) + f_-(-\infty) = f_-(\infty) \equiv 1$ is satisfied by this solution. In Fig.~(\ref{fig:fprofs}b) we plot the components of the chiral density matrix (\ref{rhocomp}) for the same solution. The imaginary part of the $\rho_{\rm LR}$ goes to zero when $z\rightarrow -\infty$, as a result of our choice that $m$ becomes asymptotically real in the broken phase. The fact that the asymptotic state is an eigenstate of the effective Hamiltonian in the broken phase is seen in that $\rho$ tends to a constant.  Diagonal components of $\rho$ become large in the broken phase. This is easy to understand, because $Tr[\rho] = \langle g_0^sÊ\rangle$ is normalized to represent particle density (per unit energy and $k_\pp$-volume in the phase space, see footnote 6). However, since flux is conserved, a small local velocity $v_z = k_z/\omega$ enhances density in the broken phase by a factor of $1/v_z$. 
\begin{figure}
\centering
\includegraphics[width=7cm]{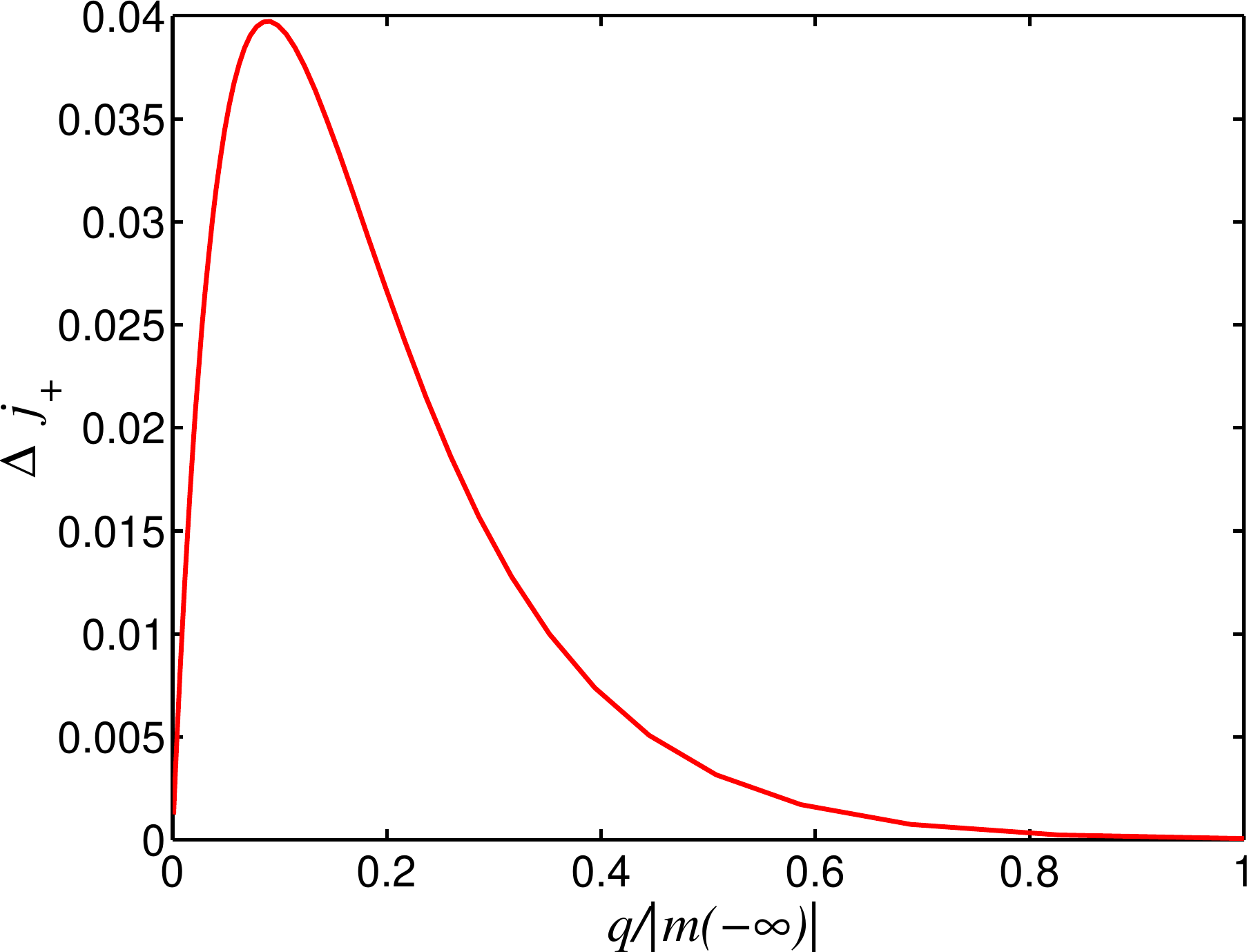}
    \caption{Shown is the current asymmetry of reflected
             states as a function of the asymptotic momentum to the mass
             ratio in the broken phase. For the wall width and the total
             change of the phase we used $\ell_w = 2$ and $\Delta\theta=-1$.}
    \label{fig:deltaj}
\end{figure}

In figure \ref{fig:deltaj} we show the particle-antiparticle flux-asymmetry $\Delta j_+ \equiv (f_+ - \bar f_+)_{z=\infty}$ as a function of  $q/|m_{-\infty}|$ where $q \equiv \sqrt{\omega^2 - |m_{-\infty}|^2}$.
In order to compute the density matrix for antiparticles we simply need to make the transformation $m\rightarrow m^*$. The characteristic peaked shape of the flux-asymmetry arises as follows: for $\omega < |m_{-\infty}|$ particles cannot enter the broken phase at all, resulting in a complete reflection both in particle and in antiparticle sectors. When $q$ is positive, the reflection amplitudes and hence the reflected asymmetry start to grow.  This growth is cut for large $q$ when the particle compton wavelength becomes shorter than the wall width and the particles start to behave classically, and the reflection amplitudes and hence also the asymmetry start to decline again (exponentially). These results were derived earlier for example ref.~\cite{CKV} using the Dirac equation approach.  We considered this case here partly to show how our formalism works in a nontrivial situation; in particular one sees again the necessity of including the $k_z=0$-solutions into the definition of the full density matrix. Second, this application is precisely the one we wish to solve in completeness, including the collisions, in the EWBG problem~\cite{HKR}.

\subsection{Homogenous case, particle number in the early universe}

We introduced the concept of the weighted density matrix in the case of static planar symmetric problem. The notion is of course more general and equally well adaptable to the homogenous problems. One interesting application of our formalism concerns the definition of the particle number in the early universe. This problem was recently considered by Garbrecht \etal in \cite{TomGar}. We shall now see how these results follow straightforwardly in the present formalism. Extension of this analysis to the case with interactions will be published elsewhere~\cite{HKR2}.

\begin{figure}
\centering
\includegraphics[width=14cm]{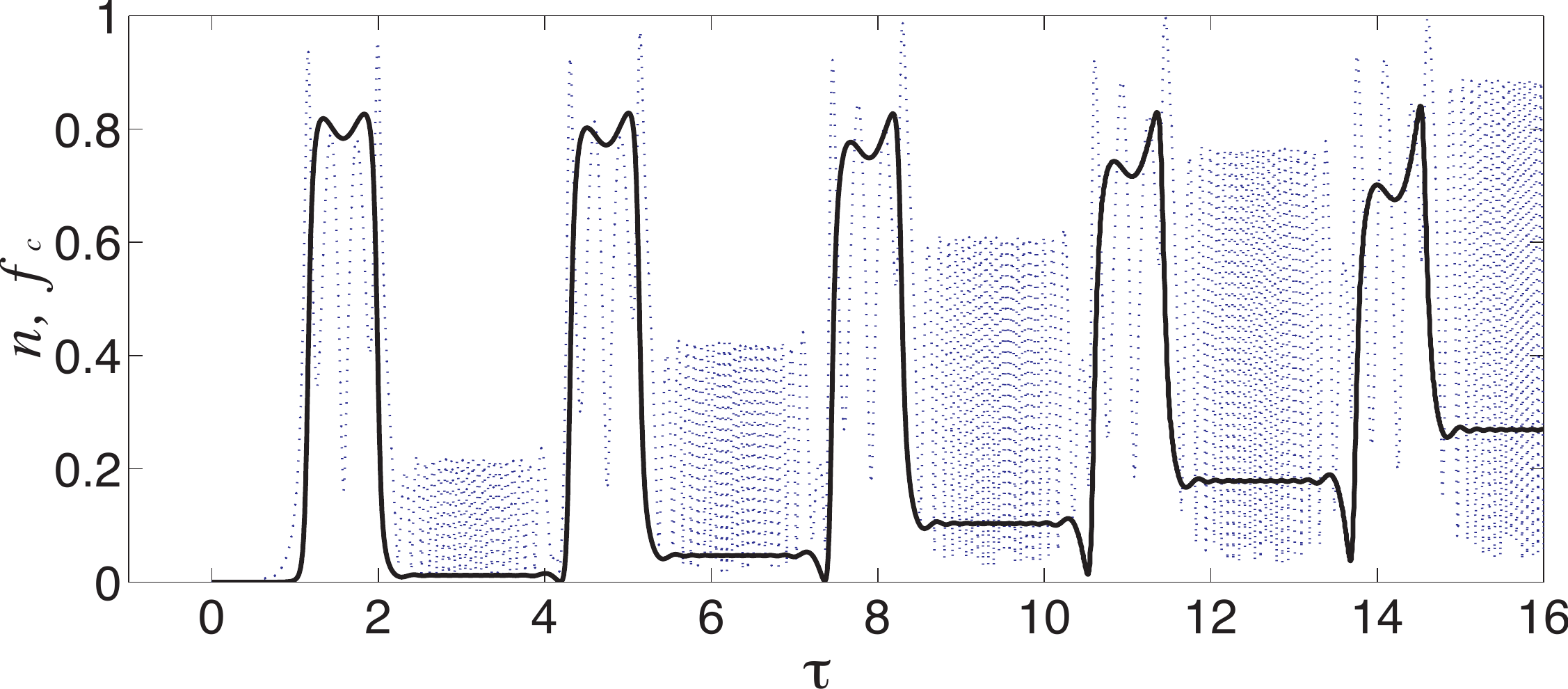}
\caption{Shown is the fermion number density (thick line) $n_{{\vec k} h}$ as a function of time  $\tau = \omega_\phi t$ for negative helicity $h = -1$ in an inflatory preheating model defined in ref.~\cite{TomGar}. For antifermions $\bar n_{{\vec k} h} = n_{{\vec k} h}$.  Thin dotted line shows the overall amount of coherence between fermions and antifermions defined as $f_c \equiv \sqrt{f_{h1}^2+f_{h2}^2}$.  Effects of inflaton oscillation are modelled by a varying mass term $m(t) = (10 + 15 \cos(2 \omega_\phi t) + i \sin(2 \omega_\phi t)) |\vec k |$ where $\omega_{\phi}= |\vec k |$ is the frequency of the oscillation.}
\label{Fig:partnumb}
\end{figure}

In the homogenous case the spatial momenta $\vec k$ and the helicity are good quantum numbers. Since the system is also isotropic, it is natural to consider density matrices of the form:
\beq
  \rho_{\cal W} ( k_0,  |\vec k|,  h; t)
     \equiv \sum_{h'} \int \frac{{\rm d}k'_0}{2\pi} \frac{{\rm d}^3
       k'}{(2\pi)^3}\; {\cal W} (k_0, |\vec k|, h \mid\hskip-0.5truemm\mid k'_0,|\vec k'|,h'\,;\,t)
      \; g^<_{h'}(k'_0,|\vec k'|;t) \,.
\label{rhoWh}
\eeq
The physical situation considered in ref.~\cite{TomGar} is that of a coherent particle production in the early universe by an oscillating homogenous scalar field coupled to fermions. Unlike momentum and helicity, energy is not conserved and, having no prior information on it, we have to leave energy completely unspecified in the definition of the physical density matrix. Then the appropriate weight function is:
\beq
{\cal W}_1 =  \frac{(2\pi)^3}{4\pi |\vec k'|^2} \delta(|\vec k|-|\vec k'|) \, \delta_{h,h'} \, ,
\label{weight2h}
\eeq
where the factor in front of deltas is introduced to normalize the 3-dimensional phase space density of states.  It is now easy to see that the corresponding weighted density matrix
\beqa
\rho_{{\cal W}_1}(k_0,|\vec k|,h;t) &=& \sum_{h'} \int 
\frac{{\rm d}k'_0}{2\pi} \frac{{\rm d}^3k'}{(2\pi)^3}\;
\frac{(2\pi)^3}{4\pi |\vec k'|^2} \delta(|\vec k|-|\vec k'|) \, \delta_{h,h'}
 \; g^<_{h'}(k'_0,|\vec k'|;t) \nonumber\\ 
&=& \int \frac{{\rm d}k_0}{2\pi} \; g^<_{h}(k_0,|\vec k|;t) \equiv \;  \rho_h(|\vec k|;t)
\label{smearedrhoext}
\eeqa
obeys the evolution equation:
\beq
\partial_t \rho_h = -i[H, \rho_h] \,,
\label{rho2}
\eeq
which now is perfectly sensible, nonsingular equation where $H$ is given by Eq.~(\ref{Hamilton}). If one introduces the Bloch-representation for the weighted density matrix:
\begin{equation}
\rho_h \equiv \frac{1}{2}(\langle g^h_0 \rangle + \langle \vec{g}^h \rangle \cdot \vec{\sigma})\,,
\end{equation}
one can write the equation (\ref{rho2}) for the integrated components $\langle g^h_\alpha \rangle \equiv \int \frac{{\rm d}k_0}{2\pi} \, g^h_\alpha$  as
\beqa
\partial_t \langle g^h_0\rangle	 &=& 0
\nonumber \\
\partial_t \langle g^h_1\rangle &=&  2h|\vec{k}| \langle g^h_2\rangle - Ê2m_I \langle g^h_3\rangle 
\nonumber \\
\partial_t \langle g^h_2\rangle &=& - 2h|\vec{k}| \langle g^h_1\rangle - 2m_R \langle g^h_3\rangle  
\nonumber \\
\partial_t \langle g^h_3\rangle &=&  2 m_R \langle g^h_2 \rangle + 2 Êm_I \langle g^h_1\rangle \,,
\label{AHset1}
\eeqa
This equation is equivalent to Eqs.~(31-40) in ref.~\cite{TomGar}\footnote{Note that the signs of our functions $\langle g_{2,3}^h\rangle$ differ from the corresponding ones in~\cite{TomGar} because we define our 2-point functions differently (see footnote 1).}.  The authors of ref.~\cite{TomGar} did not consider the constraint equations in their work, and averages $\langle g_{\alpha}^h\rangle$ were introduced as unspecified moment functions, whose connection to the particle number had to be worked out using operator formalism and Bogolybov transformations. In our treatment the mass-shell functions $f^h_{\pm}$ are directly related to the desired particle number densities. Indeed, for the fermionic current density we get:   
\begin{equation}
\langle \bar\psi \gamma^0 \psi \rangle 
 = \int \frac{{\rm d}^4k}{(2\pi )^4} {\rm Tr}[\bar G^< ] 
 = \sum_h \int \frac{{\rm d}^3k}{(2\pi )^3} \langle g^h_0 \rangle
 =  \sum_h \int \frac{{\rm d}^3k}{(2\pi )^3} (f^h_{+} + f^h_{-})Ê\,. 
\end{equation}
In the last step we used the explicit form of $\langle g^h_0 \rangle$ in terms of $f^h_{\pm}$'s; the complete set of expressions for all components is:
\begin{eqnarray}
\langle g^h_0 \rangle &=& f^h_{+} + f^h_{-}
\nonumber \\[3mm]
\langle g^h_1 \rangle &=& \frac{m_R}{\omega}(f^h_{+} - f^h_{-}) + f^h_{1}
\nonumber \\[1mm]
\langle g^h_2 \rangle &=& -\frac{m_I}{\omega}(f^h_{+} - f^h_{-}) + f^h_{2}
\nonumber \\
\langle g^h_3 \rangle &=& -h\frac{|\vec k|}{\omega}( f^h_{+} - f^h_{-}) + h \Big(
\frac{m_R}{|\vec k|}f^h_{1} - \frac{m_I}{|\vec k|}f^h_{2}\Big).
\label{rhocompbloch}
\end{eqnarray}
According to the Feynman-Stueckelberg interpretation the actual phase-space particle number densities are $n_{\vec{k}h} \equiv f^h_{+}(|\vec k|)$ for fermions and $\bar n_{\vec{k}h} \equiv 1 - f^{h < }_{-}(|\vec k|)$ for antifermions.  Thus, using the inverse relations of Eqs.~(\ref{rhocompbloch}) we get for a given 3-momentum $\vec{k}$ and helicity $h$:
\beq
n_{\vec{k}h} \equiv \frac{1}{2 \omega}\left(-h|\vec k| \langle
Êg^h_3 \rangle + m_R \langle g^h_1 \rangle - m_I \langle g^h_2
Ê\rangle \right) + \frac{1}{2} \langle g^h_0 \rangle 
\label{partnumber}
\eeq
By setting a constraint ${\rm Tr}[\rho^h] = \langle g^h_0 \rangle \equiv 1$ this reduces to the expression used to define the particle number in~\cite{TomGar}, apart from some sign conventions (see footnote 7).  Similarly, the antiparticle number is found to be
\begin{equation}
{\bar n}_{\vec{k}h} = n_{\vec{k}h} - \langle g^h_0 \rangle + 1 \,.
\end{equation} 
Setting $\langle g^h_0 \rangle \equiv 1$ thus corresponds to assuming zero chemical potential: $\langle g^h_0 \rangle \equiv 1 \Rightarrow n_{\vec{k}h} = {\bar n}_{\vec{k}h}$.  In Fig.~\ref{Fig:partnumb} we plot the particle number $n_{\vec{k}h} = f_{h+}(|\vec k|)$ and a function $f_c \equiv (f_{h1}^2+f_{h2}^2)^{1/2}$, which measures the overall coherence between particles and antiparticles, in the case of a time dependent mass term, corresponding to an oscillating inflaton field during inflatonary preheating, introduced in ref.~\cite{TomGar}. We note that the generation of the particle number is highly coherent phenomenon with the amplitude of quantum coherence increasing with each oscillation period of the inflaton field.  In ref.~\cite{HKR2} we generalize our present formalism to the case with interactions and show how the interactions change the particle number production and how they introduce the quantum decoherence leading to eventual statistical ensemble of particles. 

\section{Generalization to flavour mixing}
\label{sect:flavourmixing}

In this section we generalize our results to a case with several different flavours, \ie when the mass function is replaced by an $N\times N$-matrix $M$. 
First note that this generalization has nothing to do with the chiral decompositions we made in arriving equations (\ref{Hermitian22}-\ref{AntiHermitian22}) in the time-dependent case or to equations (\ref{Hermitian22z}-\ref{AntiHermitian22z}) in the case with planar symmetry. The only difference is that the mass operators in the generalized Hamiltonian (\ref{GenHamiltonian}) and momentum operators (\ref{Pamilton}) are to be replaced with matrix operators.
\beq
i \partial_t g^<_h = \hat H g^<_h - g^<_h \hat H^\dagger \,, \qquad
2k_0 g^<_h = \hat H g^<_h + g^<_h \hat H^\dagger \,,
\label{rhoHOMOGNN}
\eeq
with %
\beq
\hat H = \left( \begin{array}{cc}
                   -h|\vec k| & \hat M \\
                   \hat M^\dagger & h|\vec k|
                 \end{array}\right) \,,
\label{HamiltonNN}
\eeq
in the homogenous case and
\beq
is \partial_z g^<_s = \hat P g_s^< - g_s^s \hat P^\dagger \,, \qquad
-2 s k_z g^<_s = \hat P g_s^< + g_s^< \hat P^\dagger \,,
\label{rhozNN}
\eeq
with
\beq
\hat P   \equiv \left( \begin{array}{cc}
                   k_0 & -\hat M \\
                   \hat M^\dagger & -k_0
                 \end{array}\right)
\label{PeqNN}
\eeq
in the planar symmetric case. Here we define mass operators 
$\hat M \equiv M\exp{(-\frac{i}{2}\partial_x^M \cdot \partial_k)}$ and 
$\hat M^\dagger \equiv M^\dagger\exp{(-\frac{i}{2}\partial_x^M \cdot \partial_k)}$, where $\partial_x^M$-derivatives always operate on the mass matrices, and $\partial_k$-derivatives on matrix functions $g^<_{s,h}$. In the mean field limit the Hamiltonian operator $\hat H$ becomes a Hermitian local matrix operator $H$ and the right hand sides of the kinetic and constraint equations in (\ref{rhoHOMOGNN}) become a commutator $[H,g^<_h]$ and an anticommutator  $\{H,g^<_h\}$, respectively. (Note that $(\hat M)^\dagger \neq \hat M^\dagger$, except in the mean field limit.) Equations (\ref{rhoHOMOGNN}-\ref{HamiltonNN}) and (\ref{rhozNN}-\ref{PeqNN}) are completely general in the collisionless limit. However, their interpretation is complicated due to the same issues related to the singular spectral shell solutions to the constraint equations we found in the case of a scalar mass function. In what follows, we shall consider  the effects of flavour mixing in the planar symmetric case.  

\subsection{Planar symmetric case with flavour mixing}

As before, we shall constrain our analysis to the mean field limit $\hat M \rightarrow M$. Moreover, for the simplicity of notation, we shall take $M$ to be Hermitian $M = M^\dagger$. Generalization to a non-restricted complex mass matrix is straightforward, but the non-Hermitian structure is not relevant for the qualitative issues we wish to discuss here.  Introducing again a Bloch representation in chirality, we can write the constraint equations in (\ref{rhozNN}) in a component form
\beqa
k_0 g^s_0 + sk_z g^s_3 - \frac{1}{2} \{M, g^s_1\} &=& 0
\nonumber \\
k_0 g^s_3 + sk_z g^s_0 - \frac{i}{2} [M, g^s_2] &=& 0
\nonumber \\
sk_z g^s_2 - \frac{i}{2} [M, g^s_0] &=& 0
\nonumber \\
sk_z g^s_1 + \frac{1}{2} \{M, g^s_3\} &=& 0 \,,
\label{Cset2}
\eeqa
where $M$ and $g^s_\alpha$ are NxN matrices in the flavour space. Let us again first study the case where $k_z \neq 0$. From the three last equations in (\ref{Cset2}) one obtains
\beqa
g^s_2 &=& \frac{is}{2k_z}[M,g^s_0] 
\nonumber \\
g^s_3 &=& -\frac{sk_z}{k_0}g^s_0 - \frac{s}{4k_zk_0}\left[M,[M,g^s_0]\right] 
\nonumber \\
g^s_1 &=& \frac{1}{2k_0}\{M,g^s_0\} 
 + \frac{1}{8k_z^2k_0}\left\{ M,\left[M,[M,g^s_0]\right]\right\}.
\label{chsoltns}
\eeqa
Putting these solutions back to the first equation in (\ref{Cset2})
then gives the spectral equation:
\beq
(k_0^2 - k_z^2 - M^2) g^s_0 + \frac{1}{2}[M^2,g^s_0] 
- \frac{1}{16k_z^2}\left[ M^2,[M^2,g^s_0] \right] = 0.
\label{SpecEq2}
\eeq
In order to carry the analysis further, we need to go to the basis where the mass matrix is diagonal. Since $M$ is assumed to be Hermitian, there is a unitary matrix $U$ such that
\beq 
m_D \equiv U M U^\dagger
\label{mD}
\eeq
is diagonal. Correspondingly, we denote the density matrix in the diagonal basis by: 
\beq
g^s_{D\alpha} \equiv  U g^s U^\dagger \,.
\eeq
The first  observation to be made is that the two last terms in the equation 
(\ref{SpecEq2}) are purely off-diagonal in the mass eigenbasis:
\beqa
  [m_D^2,g^s_{D0}]_{ij} &=& - \Delta m^2_{ij} (g^s_{D0})_{ij}
 \nonumber \\
  \left[ m_D^2, [m_D^2,g^s_{D0}]\right]_{ij} &=& (\Delta m^2_{ij})^2 
  (g^s_{D0})_{ij},
\label{commutators}
\eeqa
where $\Delta m^2_{ij} \equiv m_{Dj}^2 - m_{Di}^2$. Using the results 
(\ref{commutators}), we can write (\ref{SpecEq2}) in the component form:
\beq
\left( k_0^2 - k_z^2 - M_{ij}^2 - \frac{(\Delta m^2_{ij})^2}{16k_z^2} \right) 
   (g^s_{D0})_{ij} = 0.
\label{SpecEq2b}
\eeq
where $M^2_{ij} \equiv (m_i^2+m_j^2)/2$~\footnote{This dispersion relation was found in ref.~\cite{Tomislavitaas}, but the physical content of the small $k_z$-branch as one corresponding to the quantum coherence was not realized by the the authors of that paper.}. Let us first consider the diagonal entries in the equation (\ref{SpecEq2b}). Because $\Delta m^2_{ii} =0$ it is immediately clear that diagonal equations have solutions analogous to the solutions found in the case with a scalar mass function. The dispersion relation is
\beq
k_{zii} = \pm \sqrt{k_0^2 - m_{Di}^2}
\label{DR2d}
\eeq
and the corresponding spectral solution for the density matrix elements:
\beq
(g^s_{D0})_{ii} =
             2 \pi  f^s_{ii, s_{k_z}}(k_0,z) \, \frac{|k_0|}{|k_z|} 
             \,\delta(k_z - s_{k_z} \sqrt{k_0^2 - m_{Di}^2})\,.
\label{SpecSol2d}
\eeq
Moreover, it is easy to see that for the diagonal elements the equations  (\ref{chsoltns}) reduce to equations (\ref{g13}) and similarly that the  diagonal parts of the evolution equations reduce to the equivalent expressions in the scalar mass case. We then obtain the following solutions for the chiral structure of the diagonal matrix elements:
\beq
(g^<_{sD})_{ii}  = \frac{1}{2} (g^s_{D0})_{ii} \left( \begin{array}{cc}
                                  1 - sk_z/k_{0ii}  &  m_{Di}/k_{0ii} \\
                                  m_{Di}/k_{0ii}       &  1 + sk_z/k_{0ii} 
                                 \end{array} \right) ,
\label{simpleqs2}
\eeq
which, as before, is seen to describe free propagation of a given helicity mass eigenstate.  Before moving on to discuss the off-diagonal constraint equations on mass-shells, let us now find out if any $k_z=0$-solutions might be left out by our previous derivation.  Setting $k_z=0$, the constraints become 
\beqa
2k_0 g^s_{D0} - \{ m_D, g^s_{D1} \} &=& 0
\nonumber \\
2k_0 g^s_{D3} - i [m_D, g^s_{D2}] &=& 0
\nonumber \\
\phantom{a }[m_D, g^s_{D0}] &=& 0
\nonumber \\
\{ m_D, g^s_{D3} \} &=& 0 \,.
\label{Cset4}
\eeqa
The last of these equations immediately implies that $g^s_3(k_z=0)=0$. The
commutator constraint on the third line on the other hand only sets
$g^s_{D0_{i\neq j}} = 0$ at $k_z=0$, but leaves the diagonal elements 
of $g^s_{D0}$ arbitrary. Taking commutators and anticommutators of first 
and second equations, respectively, with respect to $m_D$ similarly leads to 
constraints $g^s_{D1,2_{i\neq j}} = 0$, but again leaves 
$g^s_{D1,2_{ii}}$ arbitrary. Restricting now to diagonal part of the 
first constraint on (\ref{Cset4}), we easily find
\beq
g^s_{D0_{ii}} = \frac{m_{Di}}{k_0} g^s_{D1_{ii}},
\label{diagonalk0}
\eeq
at $k_z=0$. That is, we find that all off-diagonal components 
of $g^s_{D\alpha}$ must vanish on shell $k_z=0$. Moreover, also the 
diagonal elements of $g^s_{D3}$ vanish while the components
$g^s_{D1,2_{ii}}$ remain arbitrary and the nonzero diagonal elements of
$g^s_{D0}$ are given by (\ref{diagonalk0}). From our previous results it
is obvious that {\em these solutions encode the information of the quantum 
coherence between the mass eigenstates of opposite helicity and momentum.}
The different on-shell solutions with their interpretations for the 
mass eigenstates are shown in the figure \ref{fig:drel} for
the case of $2\times 2$-flavour mixing.

Let us now turn to the off-diagonal dispersion relations in 
Eq.~(\ref{SpecEq2b}). These solutions describe the quantum coherence 
between {\em different} mass eigenstates. The most striking feature
about the off-diagonal dispersion relation is that for large $k_0$ it has
{\em two} distinct solutions; one for $k_z$ close to the diagonal shell
momentum and another one close to $k_z=0$. Moreover, no solutions for
the dispersion relation exist for $k_0 < {\rm max} (m_{Di},m_{Dj})$.
The interpretation of these solutions is easier when we rewrite the
dispersion relation (\ref{SpecEq2b}) in a different form:
\beq
(k_z)_{ij} = \pm \frac{1}{2}\left(\sqrt{k_0^2-m_{Di}^2} 
                              \pm \sqrt{k_0^2-m_{Dj}^2}\right) \,.
\label{speceq3}
\eeq
In this form we now explicitly indicate also that the momentum shell depends on the off-diagonal entry in question. The first signs in Eq.~(\ref{speceq3}) refer just to the direction of the momentum, while  the second two signs refer to the high- and low-$k_z$ branches of each continuous dispersion curve, to the right and to the left from the minimum  set by $k_0 = {\rm max} (m_{Di},m_{Dj})$ respectively. This structure is depicted in Fig.~\ref{fig:drel} in the 2x2-mixing case. The off-diagonal momentum shells thus correspond to the mean momenta of the mixing mass eigenstates. We can now make the following physical interpretations:

\begin{figure}
\centering
\includegraphics[height=10cm]{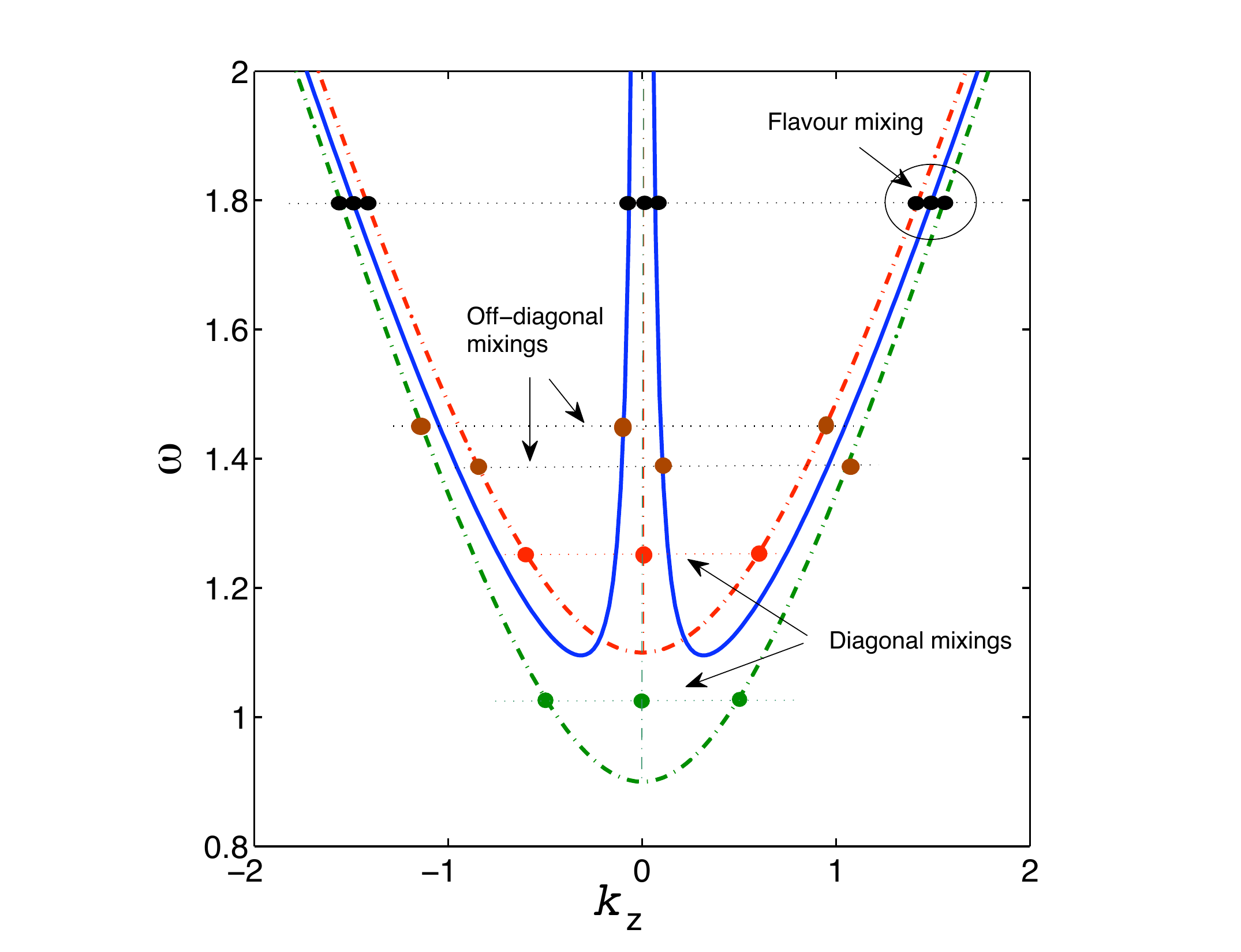}
    \hskip -4cm
    \caption{The complete set of dispersion relations corresponding to Eqs.~(\ref{SpecEq2b}) or (\ref{speceq3}).}
    \label{fig:drel}
\end{figure}

\begin{itemize}
\item{} The large momentum solutions (with plus sign inside the 
parenthesis in Eq.(\ref{speceq3}) carry the information of the 
quantum coherence related to flavour mixing between different mass 
eigenstates of {\em equal helicities} moving in the {\em same 
direction}. 
\end{itemize}

\noindent
This is of course just the quantum coherence phenomenon that is relevant 
for example for neutrino-oscillations. A triplet of such potentially mixing states with positive $k_z$ is shown by a circle in Fig.~\ref{fig:drel}. 
Second, 

\begin{itemize}
\item{} The small momentum solutions (with minus sign inside the parenthesis 
in Eq.(\ref{speceq3}) carry the information of the quantum coherence 
related to flavour mixing between different mass eigenstates of 
{\em opposite helicities} moving in {\em opposite directions}. 
\end{itemize}

\noindent These solutions are the analog of the $k_z=0$-solutions found to host the quantum coherence relevant for the reflection problem in the case of a scalar mass and for the diagonal entries in the case of a mass matrix. At this point one might appreciate the fact that we did not obtain any $k_z=0$-shell contributions for off-diagonal terms above; these solutions were already included in the ``on-shell" dispersion relations (\ref{SpecEq2b}). In fact, 
Eq.~(\ref{speceq3}) actually contains also the $k_z=0$-solution for the 
diagonal entries, as this is what the small-$k_z$ solution collapses to 
when the masses $m_{Di}$ and  $m_{Dj}$ are equal. This solution is simply 
hidden in the form (\ref{SpecEq2b}) of the dispersion relation when one first 
takes the limit $\Delta m^2_{ij}\rightarrow 0$. In Fig.~\ref{fig:drel} we show the complete set of dispersion relations for this problem.  The thick dash-dotted red and green parabolas show the diagonal mass-shell dispersion relations and thin dash-dotted lines at $k_z=0$ show the coherence shells corresponding to the diagonal solutions. Triplets of (red and green) dots on the dash-dotted lines show  particular sets of phase space elements involved in the potential diagonal mixing. The thick solid blue line shows the off-diagonal dispersion relation and the triplets of (brown) dots show shells that are involved in the off-diagonal flavour mixing between states moving to opposite directions. Finally, the set of three black dots enclosed in a circle show a triplet of shells involved in the usual flavour mixing between states moving to a same direction. We have used different values of $k_0$ in different cases just for the sake of clarity of presentation. In reality, in a case with no information on $k_z$ (like in the reflection cases we considered in sections \ref{kleinproblem}-\ref{sect:refl}) there are nine different kinematical shells with sixteen unknown functions, contributing to a physical density matrix that describes the most general mixing phenomena. This is the setting one expects to encounter (but with a generic, non-Hermitian mass matrix) in the case of a chargino reflection off a phase transition wall in an application of our formalism to the Electroweak baryogenesis.

\subsubsection{Evolution equation for a neutrino beam}

Having now explained the physical significance of different shells, we move on to briefly discuss the dynamical equations in the case of a free propagation. In the case of reflection problem the new aspect is the need to account for the mixing, as induced by locally varying unitary operator $U(z)$. We shall not consider this problem in its full generality in this paper. Instead, let us now ignore the small $k_z$-branches entirely and concentrate on flavour mixing between states of (nearly) equal momentum.  That is, we are assuming that we have a good resolution of the direction and the magnitude of the momentum, but not good enough to separate the different flavour mixing shells from each other. For simplicity we will assume perfect information of $k_0$ here. This is of course just the case of interest in case of a neutrino beam, or a neutrino flux moving in a given direction, such as reactor or solar neutrinos. Now, the matrix evolution equations written in the component form for the physcial density matrix are:
\beqa
s \partial_z \langle g^s_3 \rangle - i[M, \langle g^s_1 \rangle ] &=& 0
\nonumber \\
s \partial_z \langle g^s_0 \rangle + \{M, \langle g^s_2\rangle \} &=& 0
\nonumber \\
s\partial_z \langle g^s_2 \rangle - 2k_0 \langle g^s_1 \rangle + \{M, \langle g^s_0\rangle \} &=& 0
\nonumber \\
s\partial_z \langle g^s_1 \rangle + 2k_0 \langle g^s_2 \rangle + i [M,\langle g^s_3\rangle ] &=& 0.
\label{Kset2}
\eeqa
Using the integrated form of the third equation in (\ref{Cset2}) we can eliminate $\langle g^s_2\rangle$ from the equation for $\langle g^s_0\rangle$, which becomes simply
\beq
i \partial_z \langle g^s_0 \rangle 
   =  [M^2,\langle \frac{g^s_0}{2k_z} \rangle]
\simeq \frac{1}{2k} [M^2,\langle g^s_0 \rangle] ,
\label{evoeqflav}
\eeq
where $k$ corresponds to the mean momentum as given by the observational accuracy formally encoded in a weight function ${\cal W}$ for the problem. It is in fact easy to verify that all components $\langle g^s_\alpha\rangle$ obey an identical evolution equation, such that (\ref{evoeqflav}) actually describes the flavour evolution of a freely propagating state of a given definite spin and momentum, but with explicit flavour mixing. Of course the spin, which no more shows explicitly in Eq.(\ref{evoeqflav}) can be replaced by helicity, by associating the $z$-axis with the direction of motion of the particle.

Let us make a couple of remarks on the solution (\ref{evoeqflav}). First, it is
just the analog of the free particle motion found in sections 4-5; either for a scalar mass function, or more generally for all diagonal elements the commutator in Eq.~(\ref{evoeqflav}) vanishes.  Second, we can interpret the equation (\ref{evoeqflav}) as arising from a Hamiltonian form:

\beq
i \partial_z \rho = [H,\rho] \,,
\label{rhoeq1}
\eeq
if one makes the formal identification  $H = \sqrt{k^2+M^2} \simeq k + M^2/2k$.
In this case equations (\ref{evoeqflav}) and (\ref{rhoeq1}) are identical to first order in $M^2/k^2$.  Note however that the equation (\ref{evoeqflav}) is in fact {\em exact} (apart from the averaging in the last step) and its definition exactly encodes the amount of the information (momentum resolution) on the system. It should be obvious, that at the limit when the resolution encoded by ${\cal W}$ becomes accurate enough to separate different shells in the flavour mixing triplets shown in Fig.~\ref{fig:drel}, the density matrix becomes trivial and the commutator term vanishes in Eq.~(\ref{evoeqflav}), and the solution is reduced to a free propagation of a fixed mass, momentum and helicity eigenstate.  Note that our formalism allows one to consider also the intermediate cases where one has {\em partial} information on the flavour content of the state, such that this information affects, but does not stop completely the mixing and oscillation pattern.

\section{Interacting fields}
\label{sect:interaction}

The goal of this paper was to set up the density matrix formalism for treating quantum coherence phenomena in classical backgrounds. Eventually we wish to extend these methods to include cases with collisions. We will not pursue this goal further here, apart from a qualitative discussion of the elements needed in the derivation. For the time dependent case the generalization is actually quite straightforward, and we will present our complete results in a companion paper~\cite{HKR2}.  Most results found in this paper were derived from the free collisionless equation (\ref{complexgreen}) for the correlation function $iG_{\cal C}(u,v) = \left\langle T_{\cal C}\left[\psi(u) \bar \psi (v)\right]
\right\rangle $, defined on a complex Keldysh time-path shown in Fig.~\ref{fig:keldysphath}. More generally, in the presence of interactions, $G_{\cal C}(x,y)$ obeys the contour Schwinger-Dyson equation:
\beq
G_{\cal C} (u,v) = G^0_{\cal C} (u,v)
            + \int_{\cal C} {\rm d}^4z_1 \int_{\cal C}
                            {\rm d}^4z_2 \; G^0_{\cal C} (u,z_1)
              \Sigma_{\cal C} (z_1,z_2) G_{\cal C} (z_2,v)\,,
\label{SD1}
\eeq
where $\Sigma_{\cal C}$ is a self-energy functional, which in general depends on higher order Green's functions of the theory. This dependence eventually leads a hierarchy of coupled equations involving all possible Green's functions. The practical usefulness of the Schwinger-Dyson formalism arises from the fact that in many applications one can truncate this hierarchy to the lowest order by some reasonable approximation to $\Sigma_{\cal C}$ which only involves the 2-point functions. In the weak coupling limit for example, it is natural to do this by substituting all higher than 2-point functions by their perturbative expressions. A recent review that discusses the evaluation of $\Sigma_{\cal C}$ can be found in reference~\cite{PSW}.
\begin{figure}
\centering
\includegraphics[width=11cm]{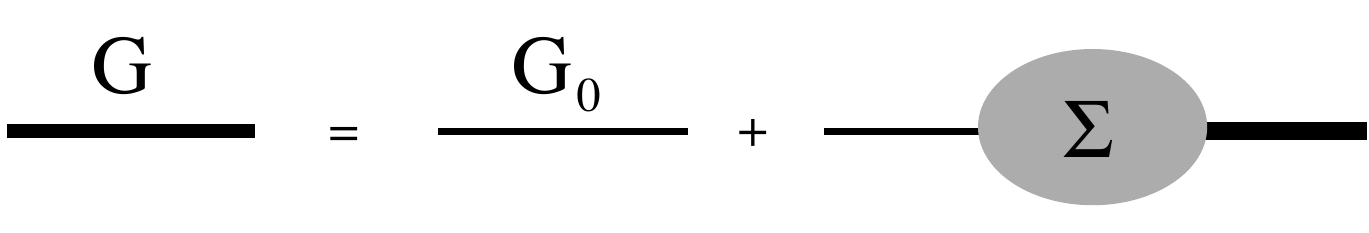}
    \caption{A generic form of a Schwinger-Keldysh equation for the
    2-point function $G^<$.}
    \label{fig:SchKelGen}
\end{figure}

Equation (\ref{SD1}) is formally expressed in Fig.~\ref{fig:SchKelGen},
where the thin lines correspond to the free particle (tree level) propagator
$G^0_{\cal C}$, and the thick lines to the full propagator $G_{\cal C}$, and 
the filled ellipsis represents the self-energy function $\Sigma_{\cal C}$.
Multiplying Eq.~(\ref{SD1}) by the inverse of the free particle propagator
$(G^0_{\cal C})^{-1}$ and integrating over the connecting variable $z_1$ one
finds
\beq
  (i\deldag_u - m^*P_L - mP_R) G_{\cal C}(u,v)
               =  \delta_{\cal C}(u^0-v^0)\delta^3(\vec u - \vec v)
               + \int_{\cal C} {\rm d}^4z
                    \Sigma_{\cal C}(u,z) G_{\cal C}(z,v),
\label{SD2}
\eeq
where $\delta_{\cal C}(u^0-v^0)$ is a contour time delta-function. Here we assumed the free Lagrangian of the form Eq.~(\ref{freeLag1}). It thus appears formally obvious how the formalism can be extended to the case with collisions; one merely needs to evaluate the appropriate function $\Sigma_{\cal C}$ and proceed in the derivation as described in this paper. There are several obstacles on the way to a set of equations that can be solved in practice however. The crucial issue turns out to be finding an approximate way to treat the phase space of the interacting system in a way that retains the notion of a single particle excitations. This can be done in a meaningful way in the so-called quasiparticle and the mean field limits (or up to first order in gradients for fermions). Taking these limits, the mixed representation equation (\ref{G-lessEq2}) becomes just
\begin{equation}
(\kdag + \frac{i}{2} \deldag_x  -  \hat m_0 - i\hat m_5\gamma^5 -\Sigma_R) G^<(x,k) = {\cal C}_{\rm coll}(x,k) \,,
\label{fulleq}
\end{equation}
where $\Sigma_R$ is the real part of the (retarded) self energy function and ${\cal C}_{\rm coll}$ is the collision integral. In the thermal equilibrium approximation ${\cal C}_{\rm coll}$ can always be written as
\begin{equation}
{\cal C}_{\rm coll} = -i \Gamma \,( G^< - G^<_{\rm eq}) \,,
\label{colleq}
\end{equation}
where $G^<_{\rm eq}$ is given by Eq.~(\ref{SK_gthermal}) and $\Gamma$ is the usual thermal collision rate. The quasiparticle approximation is familiar from thermal field theory~\cite{LeBellac}. In the present context it corresponds to neglecting all terms arising from ${\cal C}_{\rm coll}$ in the constraint equations. Under these assumptions, equations (\ref{fulleq}) can be shown to support a spectral solution for the phase space with (quasiparticle) mass and coherence shells, similar to the ones described in this paper. Given this structure to the phase space, one can define physical density matrices as weighted integrals, and compute how the collisions affect the particle distribution functions related to the various mass and coherence shells.  The resulting formalism can be used to describe for example the effects of collisions on coherent particle production in the early universe and approach to thermal equilibrium including quantitative account of the emergence of decoherence~\cite{HKR2}. 

Let us finally note that deriving an interacting theory for the static problems, is somewhat more subtle, since the usual CTP-formalism leads to Green's functions that describe correlations that vanish at temporal infinities. This is consistent with the usual definition of the asymptotically free vacuum states for the theory using temporal infinity. This is not the appropriate limit for the static reflection problems, where one rather would like to see correlations vanish at {\em spatial} infinities. Correspondingly one would like to define the vacuum states of the theory at spatial infinities, and develop a scattering formalism relating vacua and states at different spatial rather than temporal infinities~\cite{HKR}.

\section{Conclusions and outlook}
\label{sect:conclusions}

In this work we have derived quantum kinetic transport equations for fermionic systems including non-local quantum coherence. A crucial observation leading to our formalism was the finding that in cases where the full translational invariance is lost, the free fermionic 2-point correlation functions $G^{<,>}(k,X)$  have, in addition to the usual mass-shell solutions $k^2-|m|^2=0$, new solutions living on shells $k_0=0$ (homogenous case) or $k_z=0$ (planar symmetric static case). These solutions were identified as carrying the information about the quantum coherence between particle-antiparticle pairs of same helicities and opposite momenta in the former case, and between incoming and reflecting waves of equal spin in the latter case.  Another crucial element was the definition of a physical density matrix as a convolution of the singular phase space density matrix with a weight function encoding the amount of extrenous information (or the quantitative measure of the lack of it) about the state of the system. We illustrated the use of the formalism with several examples including reflection problems and definition of a particle number in the early universe during coherent particle production. We considered also the case of several mixing fermion fields and showed how the usual evolution equation for flavour mixing neutrino system arises from the singular phase space structures and an appropriate weight function. Finally we have outlined how our formalism can be extended to the case with interactions, with a slight technical reservation concerning the spatially varying problem.

Let us finally comment on our choice to limit the discussion to the mean field limit. This seems somewhat contradictory, since quantum effects become more important when the rate of change in the background gets larger in comparision with the wave length or the frequency of the probe.  Also, we got exactly the correct answers to our reflection calculations despite the mean field limit assumption. Understanding these apparent paradoxes begins from the observation that (at least for sufficiently smooth weight functions), the integrated evolution equations always have the same form as in the mean field limit, since all derivative corrections to them are reduced to vanishing surface terms. So, the only thing that {\em does} get changed by gradients is the connetion between the averaged-out density matrix elements and the mass- and coherence shell distribution functions. Indeed, our particle numbers and fluxes are just mean field approximations to the full quantum system. However, these connections do become exact in the case of the Klein problem everywhere outside the potential step, and in the case of smooth wall at spatial infinity. This is of course why the formally mean field quantities in these cases give exact results for asymptotic currents.
Similarly, in the coherent particle production case, our particle number and coherence functions provide an approximation to the full quantum phase space, that becomes exact when inflaton oscillation stops and mass becomes a constant. 

One might wonder if these considerations render our results to be only of academic interest. This is obviously not so: first the complicated structure of the phase space is not optional; it {\em is} there. One cannot just ignore the constraint equations and concentrate to the integrated form of the evolution equations. We have shown that in the mean field limit this structure is singular and allows a particularily transparent picture for separation to quantum coherence and mass-shell degrees of freedom. It is true that beyond mean field limit, the singular structure is lost. In the case of fermions this occurrs at the second order in gradients, while the first order can be computed within spectral limit and it gives rise to corrections that lead to the semiclassical effects discussed in refs.~\cite{CJK,KPSW,PSW}. Even then the mean field limit can provide a good approximation to the phase space, capturing the most important features of the quantum evolution. Second, from the practical point of view,
the singular shell structure for mass- and coherence shells is crucial when the formalism is extended to include interactions. Indeed, the entire success of the current approach relies on ones ability to find a spectral approximation to the dynamical and kinematical phase space of the system; only then can we compute the collision terms explicitly and describe the evolution of the coherence and particle numbers on these shells in a tractable manner.

As we have pointed out in many occasions, this paper is merely setting up the basic formalism which will be extended elsewhere to include decoherence~\cite{HKR,HKR2} and then applied to various problems of interest in cosmology. This formalism will be crucial in particular to reliably compute the quantum reflection contribution to the baryon number production during the electroweak phase transition~\cite{FarSha,hernandez,CKV}. It will also be possible to use it to study the effect of collisions on the coherent particle production~\cite{HKR2}. We believe that the formalism could, and also will provide to be useful in other applications beyond the immediate application to the cosmology.

\section*{Acknowledgments}  This work was partly supported by a grant from Jenny and Antti Wihuri Foundation (Herranen) and from the Magnus Ehrnrooth Foundation (Rahkila).

%
%

\end{document}